\documentclass[aps,pra,reprint,superscriptaddress]{revtex4-2}

\usepackage{graphicx}
\usepackage{amsmath,amssymb,amsfonts}
\usepackage{bm} % for bold math
\usepackage{float} % for controlling figure and table placement
\usepackage{color}
\usepackage{xcolor}
\usepackage{soul}
\setstcolor{red}
\usepackage{booktabs}
\usepackage{subcaption}
\usepackage[utf8]{inputenc}
\usepackage{multirow}
\usepackage{algorithm}
\usepackage{algorithmicx}
\usepackage{algpseudocode}
\usepackage{listings}
\usepackage{appendix}
\usepackage{caption}
\usepackage{ragged2e}
\usepackage{physics}
\makeatletter
\long\def\@makecaption#1#2{%
  \vskip\abovecaptionskip
  \noindent
  \begin{minipage}{\hsize}
    \raggedright
    #1. #2\par
  \end{minipage}
  \vskip\belowcaptionskip
}
\makeatother
\usepackage[
  colorlinks=true,
  linkcolor=blue,
  citecolor=blue,
  urlcolor=black,
  bookmarksopen=true
]{hyperref}
\usepackage[section]{placeins} % Prevent floats crossing sections

\usepackage{xcolor}
\usepackage[normalem]{ulem}
\usepackage{tikz}
\usetikzlibrary{decorations.markings}
\usetikzlibrary{calc}
\usetikzlibrary{arrows.meta, positioning}

\begin{document}

\preprint{APS/123-QED}

%\title{Resilience of subspace-based molecular excited state methods to sampling errors on quantum computers} % Force }

%\title{A Path to Quantum Utility in Chemistry: Challenging Problems and Black-Box Benchmarking of Quantum Algorithms} % Force }
\title{Chemically decisive benchmarks on the path to quantum utility}
\author{Srivathsan Poyyapakkam Sundar} 
\affiliation{Department of Chemistry, University of North Dakota, ND 58202, USA}
\author{Vibin Abraham} 
\affiliation{Physical and Computational Sciences Directorate, Pacific Northwest National Laboratory, Richland, WA, 99354,
USA.}
\author{Bo Peng$^*$} 
\email{peng398@pnnl.gov}
\affiliation{Physical and Computational Sciences Directorate, Pacific Northwest National Laboratory, Richland, WA, 99354,
USA.}
\author{Ayush Asthana$^*$} 
\email{ayush.asthana@und.edu}
\affiliation{Department of Chemistry, University of North Dakota, ND 58202, USA}
\begin{abstract}
Progress towards quantum utility in chemistry requires not only algorithmic advances, but also the identification of chemically meaningful problems whose electronic structure fundamentally challenges classical methods. Here, we introduce a curated hierarchy of chemically decisive benchmark systems designed to probe distinct regimes of electronic correlation relevant to molecular, bioinorganic, and heavy-element chemistry. Moving beyond minimal toy models, our benchmark set spans multireference bond breaking (N$_2$), high-spin transition-metal chemistry (FeS), biologically relevant iron–sulfur clusters ([2Fe–2S]), and actinide–actinide bonding (U$_2$), which exhibits extreme sensitivity to active-space choice, relativistic treatment, and correlation hierarchy even within advanced multireference frameworks.
%Together, these systems represent intermediate yet chemically realistic benchmarks that bridge the gap between minimal test cases and long-term quantum chemistry targets where quantum utility may ultimately be demonstrated.
%
As a concrete realization, we benchmark a recently developed automated and adaptive quantum algorithm based on generator-coordinate–inspired subspace expansion, ADAPT-GCIM, using a black-box workflow that integrates entropy-based active-space selection via the ActiveSpaceFinder tool. Across this chemically diverse problem set, ADAPT-GCIM achieves high accuracy in challenging correlation regimes. Equally importantly, these benchmarks expose general failure modes and design constraints—independent of any specific algorithm—highlighting the necessity of problem-aware and correlation-specific strategies for treating strongly correlated chemistry on quantum computers. To support systematic benchmarking and reproducible comparisons, the Hamiltonians for all systems studied are made openly available.
\end{abstract}

%\keywords{Suggested keywords}%Use showkeys class option if keyword
                              %display desired
\maketitle

\section{Introduction}

% Overview - key questions - problems and algorithms
Quantum computers are expected to greatly expand our capacity to solve quantum chemistry problems with the most challenging electron correlation landscapes~\cite{tilly2021variational,mcardle2020quantum,cerezo2021variational,peruzzo2014variational,magann2021pulses,kandala2017hardware,cao2019quantum,fedorov2022vqe,magann2021pulses}. 
Although significant advances have been made in hardware and algorithms, we are still far from general quantum utility on useful problems.
%While several avenues are being explored, we will focus on chemical applications of quantum computers in this article.
On computations side, major current challenges in the field include: (i) recognizing the problems to solve that are well suited for quantum computation, and (ii) developing quantum algorithms that solve them and are compatible with current hardware or the hardware of the foreseeable future.
In selecting the suitable problems, the targets should be scientifically consequential—so that their solution yields genuine advances in understanding—and exhibit electronic correlation strong enough to lie beyond the reach of state-of-the-art classical quantum chemistry methods~\cite{jiang2025walking,nutzel2025solving,alexeev2025perspective,google2025observation}, including selected configuration-interaction methods and tensor network methods~\cite{holmes2016heat,chan2011density}.
%\sout{In the design of the problem, the right problems should be important enough such that solving them makes a meaningful advance in our knowledge, and the electron correlation should be hard enough such that it is beyond all classical quantum chemistry algorithms, such as selected CI and DMRG}
On the algorithms side, the algorithms to use should be compatible with quantum computers available today or anticipated
%\sout{capable of solving these problems using quantum computers, either that are currently available or hardware that is expected} 
in the near future, spanning advanced NISQ devices and early fault-tolerant systems with the order of $10^2-10^3$ qubits~\cite{jiang2025walking,alexeev2025perspective,google2025observation}.
%\sout{(pre-fault-tolerant or fault-tolerant era quantum computers with 100-1000 qubits)}

% current state of algorithms
In the algorithms developed to solve chemical problems using quantum computers, the variational quantum eigensolvers (VQE) are currently one of the most popular algorithms~\cite{peruzzo2014variational,barkoutsos2018quantum,anand2022quantum,yung2014transistor}.
 %VQE is based on the variational principle and is a hybrid classical-quantum algorithm. It uses a quantum circuit with optimizable parameters to generate the trial state in the quantum computer. 
%Those parameters are optimized with the help of a classical computer to minimize the expectation value of the Hamiltonian with respect to the parameterized trial function on the qubits. The quality of the solution is dependent on the choice of the ansatz (parameterized quantum circuit)~\cite{peruzzo2014variational,barkoutsos2018quantum,anand2022quantum,yung2014transistor}. 
%The most commonly utilized chemistry-based ansatz for the study of molecular systems is the unitary coupled cluster singles and doubles ansatz (UCCSD) ~\cite{peruzzo2014variational,barkoutsos2018quantum,anand2022quantum,yung2014transistor}. 
However, VQE ans\"{a}tze generally have many parameters and lack guarantees of exactness, leading to challenges in achieving desired quantum advantage in strongly correlated systems. 
These limitations have motivated the development of adaptive ans\"{a}tze, such as ADAPT-VQE proposed by Grimsley et al.~\cite{grimsley2019adaptvqe}, which aim to construct more compact and problem-tailored wavefunctions. 
%In ADAPT-VQE, the ansatz is grown iteratively by selecting, at each step, the operator with the largest energy gradient, enabling exact solutions with substantially fewer parameters.
%\sout{For near-exact solutions, ADAPT-VQE by Grimsley et al.~\cite{grimsley2019adaptvqe} is a promising algorithm. It is a dynamic ansatz that grows by addition of an operator with the highest gradient at each iteration. }
Nevertheless, in the pre-fault-tolerant quantum hardware era, optimization-based algorithms, such as VQE and ADAPT-VQE, remain vulnerable to noise-induced instabilities, as the repeated evaluation of gradients and energy landscapes can be severely affected by hardware imperfections~\cite{doi:10.1021/acs.jctc.4c01657}.
%\sout{suffer because optimization in presence of noise can be unstable}
To address some of these challenges, Zheng et al. introduced the generator coordinate-inspired method, in particular its ADAPT version, the ADAPT-GCIM algorithm~\cite{qugcm,zheng2024unleashed}.
%\sout{To solve this issue, ADAPT-GCIM~\cite{zheng2024unleashed} algorithm was introduced by Zheng et al., which} 
ADAPT-GCIM dynamically builds a problem-tailored subspace by iteratively selecting operators with the largest energy gradients and diagonalizing the Hamiltonian within the resulting subspace. By shifting the focus from direct variational optimization to subspace expansion and diagonalization, ADAPT-GCIM reduces reliance on deep parameter optimization and constitutes a non-optimization-based dynamic variational quantum algorithm. In benchmark studies, ADAPT-GCIM has been shown to span a substantial portion of the relevant Hilbert space and to achieve high accuracy for a range of test systems, indicating its potential as a practical approach for quantum chemistry in the pre-fault-tolerant era. ADAPT-GCIM also reduces the simulation cost by orders of magnitude, making it easier to test on classical hardware~\cite{zheng2024unleashed}. However, like other subspace-based methods, its ultimate performance and scalability remain system-dependent and do not, by themselves, resolve all challenges associated with strongly correlated electronic structure. 
Other related diagonalization- and sampling-based approaches have also been proposed recently, including sampling quantum diagonalization (SQD)~\cite{shajan2024towards,barison2025quantum} and cyclic VQE~\cite{zhang2025cyclic}. These methods can be formulated in either optimization-based or non-optimization-based variants and typically employ shallow circuits and lower shot count scaling. Nevertheless, it remains an open question whether such approaches can consistently outperform the classical selected configuration interaction methods they closely resemble~\cite{reinholdt2025critical}.

In identifying chemical problems where quantum computers may ultimately establish a clear and defensible quantum utility, much attention has been focused on strongly correlated systems that are central to chemistry and material sciences, including metalloclusters, such as the FeMoco~\cite{montgomery2018strong}. These systems are widely regarded as flagship targets for quantum simulation due to their complex electronic structure and the limitations of existing classical methods.
At present, however, such large and strongly correlated systems remain beyond the reach of both classical exact simulators and most available quantum hardware. As a result, the majority of quantum chemistry algorithm development and benchmarking has historically relied on small molecules and minimal test cases, including hydrogen chains (H$_4$, H$_6$, etc.) and simple diatomics or triatomics and organic compounds~\cite{grimsley2019adaptive,asthana2022equation,kumar2023quantum,huggins2020non,sawaya2024hamlib}. While valuable for method validation, these systems probe only a narrow range of correlation regimes.
Recent experimental progress has slowly begun to bridge this gap. Notably, quantum simulations of chemically realistic molecules using nontrivial basis sets, such as benzene in a Dunning-type basis, have now been demonstrated on quantum hardware, marking an important step beyond minimal toy models~\cite{arxiv:2507.01199}. %These advances suggest that intermediate-scale, chemically meaningful systems may soon become accessible as benchmark problems.
In parallel, quantum advantage has also been demonstrated in adjacent domains, such as nuclear magnetic resonance dynamics simulated using quantum echo algorithms on superconducting processors~\cite{google2025observation}. %While distinct from electronic structure calculations, such results underscore the broader potential of quantum simulation for many-body quantum dynamics.
Taken together, these developments highlight the need for a more systematic and representative set of benchmark systems---intermediate in size and correlation complexity---that better reflect the classes of problems quantum chemistry algorithms are ultimately intended to address. Establishing such benchmarks would not only enable more meaningful performance comparisons, but also inform algorithm design choices, including the construction of operator pools and problem-adapted ans\"{a}tze.

The developments presented in this manuscript are organized around two complementary focus points: (i) the construction of a curated set of chemically meaningful benchmark systems spanning distinct regimes of electronic correlation, and (ii) the development of a practical black-box workflow for applying and benchmarking automated and adaptive quantum algorithms on these systems, including automated active-space selection and ans\"atze preparation.

In the first focus point, we introduce a curated set of molecular systems that pose progressively more challenging tests for quantum algorithms. These benchmarks are designed to be easily imported into Python-based workflows and treated in active spaces of varying sizes, enabling systematic assessment across qualitatively distinct correlation regimes while moving beyond traditional minimal toy models.
The simplest system in this collection is the N$_2$ molecule, which provides a well-established test of multireference bond breaking arising from its strong triple bond. Although tractable using advanced classical methods~\cite{chan2004state,liao2023density,kumar2022quantum}, N$_2$ has also served as an important benchmark for quantum algorithms, including contextual VQE approaches that exploit problem structure~\cite{weaving2025csvqe_n2}.
The next benchmark systems are the FeS molecule and the [2Fe--2S] cluster. FeS introduces a high-spin quintet ground state that exposes limitations of commonly used excitation operator pools, requiring problem-specific operator design for accurate treatment. The [2Fe--2S] cluster provides a controlled step towards metallocluster chemistry and biologically relevant targets, informing progress towards systems such as FeMoco~\cite{miyagawa2020theory,morchen2024classification}.
The final system in our benchmark collection is the actinide–actinide bonded molecule U$_2$. This system represents an extreme challenge for electronic structure theory: even the fundamental nature of the U–U bond, whether it is quadruple or quintuple, remains debated despite decades of study using state-of-the-art relativistic and multireference methods~\cite{gagliardi2005quantum,knecht2019relativistic,ciborowski2021metal}. A definitive resolution requires a rigorous treatment of relativistic effects, including spin–orbit coupling at the orbital (spinor) level, combined with a high-accuracy treatment of strong correlation in large active spaces and an accurate description of dynamical correlation. Even when restricting the active space to the valence $5f6d7s$ orbitals, the problem involves 52 spinors, corresponding to approximately 52 qubits, and larger active spaces are needed for chemically accurate ground- and excited-state properties. Our preliminary studies indicate that high-order excitations remain significant even in reduced active spaces, highlighting the exceptional difficulty of this problem and its relevance as a long-term quantum-utility target in heavy-element chemistry.

In the second focus point of this work, we present a black-box–style workflow for applying and benchmarking selected quantum algorithms with automated and adaptive features on the challenging systems introduced above. The central features of this focus point are: (i) the automated selection of chemically relevant active spaces using the entropy-based ActiveSpaceFinder tool~\cite{shirazi2025asf}, which enables consistent and unbiased comparisons across systems of varying complexity; and (ii) the construction of problem-adapted operator pools and algorithmic strategies based on the feedback from benchmarking.
As a concrete realization of this workflow, we employ ADAPT-GCIM as the primary algorithmic framework for benchmarking. Our results demonstrate that ADAPT-GCIM can achieve high accuracy for several of the most challenging systems in this collection within their respective active spaces, highlighting its potential as a powerful approach for near-term and early fault-tolerant quantum chemistry. Equally important, these studies provide concrete insights into operator pool design: different electronic configurations and spin states necessitate tailored operator sets and algorithmic strategies, underscoring the importance of problem-aware algorithm design in variational quantum computing. Further, the benchmarking gave insights into algorithmic strategies for ADAPT-GCIM for the cases where the correlation problem becomes especially complex.

The remainder of the manuscript is organized as follows. In Sec.~\ref{theory}, we provide the theoretical background for the ADAPT-GCIM algorithm and the ActiveSpaceFinder approach. In Sec.~\ref{results}, we present results obtained using noiseless classical simulators for the systems in our benchmark collection: N$_2$, FeS, [2Fe–2S], and U$_2$. Finally, in Sec.~\ref{conclusion}, we discuss the broader implications of these results for quantum algorithm development and the pursuit of quantum utility in chemistry.

%%%%%%%%%%%%%%%%%%%%%%%%%%%%%%%%%%%%%%%%%
\section{Theory}\label{theory}

%In this work, we employ the ADAPT--GCIM algorithm to obtain accurate ground-state energies for a diverse set of molecular systems spanning weakly to strongly correlated regimes. 
The ADAPT--GCIM code used here was developed in-house using the \textsc{PennyLane} framework, following the methodological principles introduced in Refs.~\citenum{zheng2024unleashed,qugcm}. To ensure a balanced, systematically improvable, and chemically meaningful treatment of electron correlation, we combine ADAPT--GCIM with the ActiveSpaceFinder (ASF) protocol of Reiher and co-workers~\cite{morchen2024classification} to determine an appropriate active orbital space for each system.
This combination is chosen to reflect a realistic near-term workflow in which chemically relevant active spaces are identified in a largely automated fashion and quantum algorithms are deployed without extensive problem-specific tuning. As such, the resulting pipeline enables systematic and reproducible benchmarking of quantum algorithms across distinct correlation regimes, providing a practical stepping stone towards quantum utility in quantum chemistry.

\subsection{ADAPT--GCIM}

The ADAPT--GCIM algorithm is a subspace-based quantum many-body method that merges three key components~\cite{qugcm,zheng2024unleashed}: (i) the gradient-guided operator selection protocol of ADAPT-VQE, (ii) the variational and computationally tractable subspace construction employing the Generator Coordinate Method (GCM), and (iii) the flexibility in switching off the parameter optimization. 

In the context of quantum chemistry benchmarking, ADAPT--GCIM is particularly attractive because it decouples variational expressiveness from deep nonlinear parameter optimization. Accuracy is achieved through systematic enrichment of a correlated subspace rather than through the optimization of a large number of variational parameters, making the method especially well suited for near-term quantum hardware where noise can destabilize optimization-based approaches.

In classical GCM~\cite{hill1953nuclear,Griffin1957collective,rodriguez2002correlations,bender2003self,ring2004nuclear,yao2010configuration,egido2016state,hizawa2021generator}, the correlated wave function is expressed as a superposition of non-orthogonal ``generator states,'' typically defined by mean-field solutions constrained along collective coordinates. Although GCM captures large-amplitude correlation effects beyond the static mean-field picture, the need for continuous generator coordinates and constrained self-consistent-field calculations leads to substantial computational overhead. The quantum GCM approach~\cite{qugcm} relaxes these requirements by generating discrete non-orthogonal states through the application of physically motivated excitation operators to a reference state, while retaining the variational structure of the Hill--Wheeler equation.

ADAPT--GCIM further enhances the quantum GCM approach by selecting the generator states adaptively. The algorithm begins from an initial reference state, usually the Hartree--Fock (HF) determinant, and employs an operator pool consisting of unitary excitation operators appropriate for the target spin and particle-number sector. At iteration $k$, the importance of each operator $\hat{O}$ in the pool is evaluated using the energy gradient
\begin{align}
    \frac{\partial E^{(k)}}{\partial \theta}
    = \big\langle \Psi_k \big| [\hat{H},\hat{O}] \big| \Psi_k \big\rangle,
\end{align}
which quantifies the coupling between $\hat{O}$ and the current variational subspace. The operator with the largest gradient magnitude is selected, and its corresponding unitary generating function,
\begin{align}
    G_n(\theta) = e^{\theta \hat{O}},
\end{align}
with a fixed, non-optimized parameter $\theta$, is used to generate new non-orthogonal basis states.

Repeated application of these generating functions yields an adaptively expanded subspace,
\begin{align} 
\{\ket{\Phi_i}\}&=\big\{\ket{\phi}, \notag \\ &~~~~~~G_1(\theta)\ket{\phi},~G_2(\theta)\ket{\phi},\cdots,~G_n(\theta)\ket{\phi},\notag \\ &~~~~~~G_2(\theta)G_1(\theta)\ket{\phi},~ \cdots,~G_n(\theta)G_{n-1}(\theta)\ket{\phi}, \notag \\ &~~~~~~\vdots \notag \\ &~~~~~~G_n(\theta)G_{n-1}(\theta)\cdots G_1(\theta)\ket{\phi} \big\}, 
\label{eq:gcimmain}
\end{align}
where $\ket{\phi}$ denotes the HF reference state. In principle, retaining all products of generators would yield an exponentially growing subspace approaching a complete configuration-interaction–like expansion. This formal limit provides a useful conceptual reference, even though the practical power of ADAPT--GCIM lies in selectively approximating this space using a small, physically informed subset of states.

In practice, ADAPT--GCIM enforces a controlled truncation of this expansion. Rather than retaining all generator products, the subspace is grown linearly by appending only a small number of states at each iteration. For a newly selected generator $G_k(\theta)$, only the following two states are added:
\begin{align}
\big\{ G_k(\theta)\ket{\phi},~G_k(\theta)G_{k-1}(\theta)\cdots G_1(\theta)\ket{\phi} \big\}.
\end{align}
This design reflects a deliberate trade-off between expressiveness and numerical stability, ensuring that subspace growth remains aligned with the dominant correlation pathways identified by the gradient criterion while avoiding exponential proliferation of basis states.

Projecting the molecular Hamiltonian $H$ onto the resulting non-orthogonal basis yields an effective Hamiltonian matrix $\mathbf{H}_k$ and overlap matrix $\mathbf{S}_k$. Solving the generalized eigenvalue problem
\begin{align}
    \mathbf{H}_k f_k = \epsilon_k \mathbf{S}_k f_k
\end{align}
provides variational estimates of the ground-state and low-lying excited-state energies. Although the generalized eigenvalue problem can, in principle, introduce numerical instabilities, thresholding techniques can be used to extract meaningful eigenvalues in most cases~\cite{kwao2025generalized,epperly2022theory}. Many subspace quantum algorithms rely on a similar structure of the working equation~\cite{motta2020determining,parrish2019quantum,huggins2020non,asthana2025quantum}. The associated many-body wavefunction is given by
\begin{align}
    \ket{\Psi_k} = \sum_i (f_k)_i \ket{\Phi_i},
\end{align}
which enables the evaluation of observables and transition properties within the adaptively constructed subspace. Notably, unlike ADAPT-VQE, ADAPT--GCIM requires no nonlinear parameter optimization; accuracy arises entirely from systematic subspace enrichment, making the approach particularly robust for noisy quantum devices and pre-fault-tolerant era of quantum computing~\cite{zheng2024unleashed}. An additional benefit of ADAPT-GCIM is also that there is no requirement of a stopping criterion in ADAPT-GCIM, unlike in ADAPT-VQE, where, since operators are repeated in ADAPT-VQE, there is no clear consensus on when to stop ADAPT-VQE iterations.

\subsection{Active Space Finder}

While ADAPT--GCIM provides a systematic route to constructing correlated many-body states, its practical efficiency depends critically on the choice of orbital basis in which the Hamiltonian is expressed. Restricting the problem to a compact yet correlation-relevant active space is therefore essential for maintaining favorable scaling and for ensuring that adaptive operator selection targets physically meaningful degrees of freedom.

To this end, we employ the ActiveSpaceFinder (ASF) to identify appropriate active orbital spaces for each system. ASF begins with an MP2 calculation to reduce the full orbital manifold to a manageable pool. Within this reduced space, a low-cost Full Configurational Interaction (FCI) or DMRG calculation is performed to evaluate single-orbital entropies, which quantify the entanglement between each orbital and the remainder of the system. Orbitals exhibiting large entropies are selected to form the active space, ensuring that the dominant electron correlation effects are captured with minimal overhead. The second-quantized Hamiltonian is then constructed in this tailored active space and passed to ADAPT--GCIM.

\subsection{Operator pools}

A central ingredient of adaptive variational algorithms such as ADAPT-VQE and ADAPT--GCIM is the choice of the operator pool from which the ansatz is constructed. The operator pool determines not only the expressiveness of the variational ansatz but also its symmetry properties, parameter efficiency, and robustness across different correlation regimes. In this work, we employ two complementary fermionic operator pools to benchmark algorithmic performance from weakly to strongly correlated systems.

In this work, we have made use of the fermionic SD (single and double excitation operators) operator pool~\cite{grimsley2019adaptvqe,Singh2025QubitEfficient}, generalized single and double operator pool, and their singlet spin versions. The SD pool serves as a natural baseline, closely mirroring classical coupled-cluster theory and performing well for weakly correlated systems. By contrast, the GSD pool contains the generalized single and double excitations, that span a larger set of correlation space. Further, the restriction of spin to singlet explicitly preserves total spin symmetry and is therefore better suited for strongly correlated and multireference regimes where spin adaptation is essential. We will discuss the singlet-generalized singles and doubles (SingletGSD) operator pool~\cite{Bertels2022SymmetryBreakingADAPT}in some detail, which should introduce the general concept of building these pools to the reader.  In GSD, the ansatz is built from a union of single- and double-excitation generators,
\begin{align}
    \{ \hat{O}_k \} = \{ \hat{O}_{j}^{i} \} \cup \{ \hat{O}_{kl}^{ij} \},
\end{align}
providing a systematic hierarchy for capturing correlation effects while maintaining a manageable operator space. Throughout this subsection, indices $i,j,k,l$ label general spatial orbitals, while $\alpha$ and $\beta$ denote the associated spin functions. The SD pool, in contrast, only contains the single and double excitation generators that are allowed on the Hartree-Fock reference. The additional operators in GSD pool allow the exploration of a larger subspace when applied to general wavefunction manifold with additional configurations beyond the Hartree-Fock reference.

Since all generated states from the SingletGSD operators satisfy $\langle \hat{S}^2 \rangle = 0$, this explicit enforcement of spin singlet symmetry mitigates unphysical spin contamination during adaptive ansatz growth. To construct the SingletGSD operator pool, we first construct the generalized singlet single-excitation operators, given by
\begin{align}
\hat{O}^{i}_{j}
= \frac{1}{\sqrt{2}}
\left(
  \hat{c}^{\dagger}_{i\alpha}\hat{c}_{j\alpha}
  + \hat{c}^{\dagger}_{i\beta}\hat{c}_{j\beta}
\right)
- \text{h.c.}, \label{eq:one-body}
\end{align}
which promote an electron from orbital $j$ to $i$ in a spin-adapted manner. These operators capture orbital relaxation and single-particle correlation effects while preserving singlet symmetry.
We then construct the generalized singlet double-excitation operators from several symmetry-distinct subclasses,
\begin{align}
\{ \hat{O}_{kl}^{ij} \}
= \{ \hat{O}_{kk}^{ii} \}
\cup \{ \hat{O}_{kk}^{ij} \}
\cup \{ \hat{O}_{kl}^{ii} \}
\cup \{ {}^{A}\hat{O}_{kl}^{ij} \}
\cup \{ {}^{B}\hat{O}_{kl}^{ij} \},
\end{align}
each of which captures a different physically relevant correlation channel and is described briefly below.
\begin{itemize}
\item 
The first class corresponds to pair excitations within the same spatial orbitals,
%
%T1 operators
\begin{align}
\hat{O}_{kk}^{ii}
= \hat{c}_{i\alpha}^{\dagger} \hat{c}_{i\beta}^{\dagger} \hat{c}_{k\beta} \hat{c}_{k\alpha}
- \hat{c}_{k\alpha}^{\dagger} \hat{c}_{k\beta}^{\dagger} \hat{c}_{i\beta} \hat{c}_{i\alpha},
\end{align}
which play an important role in describing strong on-site and pairing correlations.
\item
The second and third classes describe mixed-index double excitations,
%
%T2 operators
\begin{align}
\hat{O}_{kk}^{ij}
= \frac{1}{\sqrt{2}}
\left(
  \hat{c}_{i\alpha}^{\dagger} \hat{c}_{j\beta}^{\dagger} \hat{c}_{k\beta} \hat{c}_{k\alpha}
  + \hat{c}_{j\alpha}^{\dagger} \hat{c}_{i\beta}^{\dagger} \hat{c}_{k\beta} \hat{c}_{k\alpha}
\right)
- \text{h.c.} \label{eq:T2}
\end{align}
with $i \neq j$, and
%
%T3 operators
\begin{align}
\hat{O}_{kl}^{ii}
= \frac{1}{\sqrt{2}}
\left(
  \hat{c}_{i\alpha}^{\dagger} \hat{c}_{i\beta}^{\dagger} \hat{c}_{l\beta} \hat{c}_{k\alpha}
  + \hat{c}_{i\alpha}^{\dagger} \hat{c}_{i\beta}^{\dagger} \hat{c}_{k\beta} \hat{c}_{l\alpha}
\right)
- \text{h.c.} \label{eq:T3}
\end{align}
with $k \neq l$, which account for correlated two-electron rearrangements across different spatial orbitals.
\item
Finally, the remaining two classes,
\begin{align}
{}^{A}\hat{O}_{kl}^{ij}
&= \frac{1}{\sqrt{12}}\Big(
  2 \hat{c}_{i\alpha}^{\dagger} \hat{c}_{j\alpha}^{\dagger} \hat{c}_{l\alpha} \hat{c}_{k\alpha}
+ 2 \hat{c}_{i\beta}^{\dagger}  \hat{c}_{j\beta}^{\dagger}  \hat{c}_{l\beta} \hat{c}_{k\beta} \nonumber \\
&\quad + \hat{c}_{i\alpha}^{\dagger} \hat{c}_{j\beta}^{\dagger} \hat{c}_{l\beta} \hat{c}_{k\alpha}
+ \hat{c}_{i\beta}^{\dagger}  \hat{c}_{j\alpha}^{\dagger} \hat{c}_{l\beta} \hat{c}_{k\alpha}
- \hat{c}_{i\alpha}^{\dagger} \hat{c}_{j\alpha}^{\dagger} \hat{c}_{k\beta} \hat{c}_{l\alpha} \nonumber \\
&\quad - \hat{c}_{j\alpha}^{\dagger} \hat{c}_{i\beta}^{\dagger} \hat{c}_{l\beta} \hat{c}_{k\alpha}
+ \hat{c}_{j\alpha}^{\dagger} \hat{c}_{i\beta}^{\dagger} \hat{c}_{k\beta} \hat{c}_{l\alpha}
\Big)
- \text{h.c.}
\end{align}
with $i \neq j$ and $\ k \neq l$, and
\begin{align}
{}^{B}\hat{O}_{kl}^{ij}
= \frac{1}{2} \Big(
  \hat{c}_{i\alpha}^{\dagger} \hat{c}_{j\beta}^{\dagger} \hat{c}_{l\beta} \hat{c}_{k\alpha}
+ \hat{c}_{i\alpha}^{\dagger} \hat{c}_{j\beta}^{\dagger} \hat{c}_{k\beta} \hat{c}_{l\alpha} \nonumber \\
+ \hat{c}_{j\alpha}^{\dagger} \hat{c}_{i\beta}^{\dagger} \hat{c}_{l\beta} \hat{c}_{k\alpha}
+ \hat{c}_{j\alpha}^{\dagger} \hat{c}_{i\beta}^{\dagger} \hat{c}_{k\beta} \hat{c}_{l\alpha}
\Big)
- \text{h.c.}
\end{align}
with $i \neq j$ and $k \neq l$, capture higher-order spin-adapted correlation effects that become essential in strongly correlated and multireference regimes.
\end{itemize}

Together, these operator pools provide a flexible yet physically motivated foundation for adaptive ansatz construction. By comparing fermionic SD and SingletGSD within the same benchmarking framework, we are able to systematically assess the impact of spin adaptation, operator expressiveness, and pool design on the performance and robustness of ADAPT-GCIM across diverse quantum chemistry problems.

%%%%%%%%%%%%%%%%%%%%%%%%%%%%%%%%%%%%%%%%%
%
%Originally N2 figure was present here. 
\section{Results and Discussion}\label{results}
\begin{figure*}
    \begin{subfigure}[b]{0.48\linewidth}        \includegraphics[width=\linewidth]{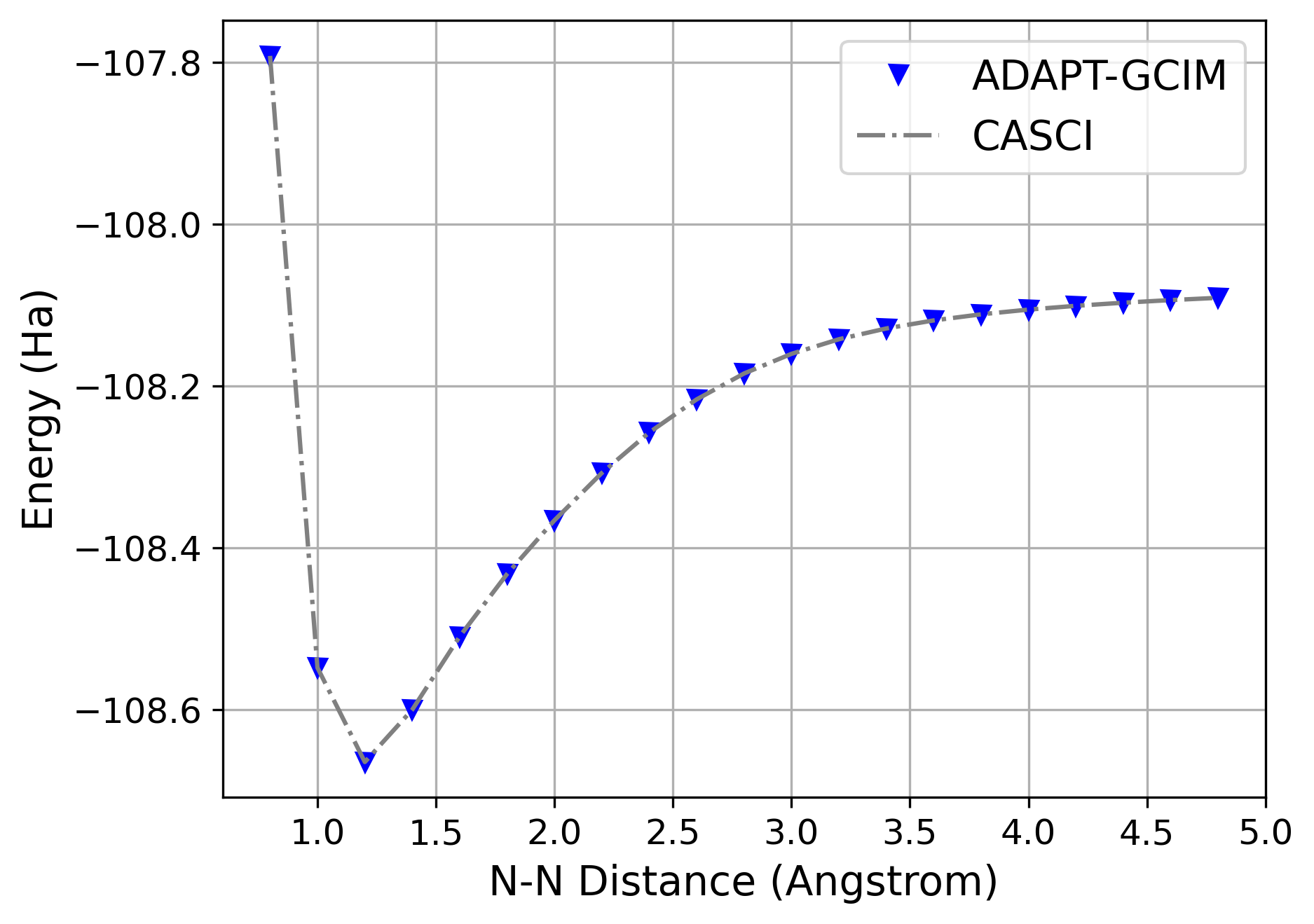}
        \caption{}
        \label{1}
    \end{subfigure}
    \hfill
    \begin{subfigure}[b]{0.48\linewidth}
        \includegraphics[width=\linewidth]{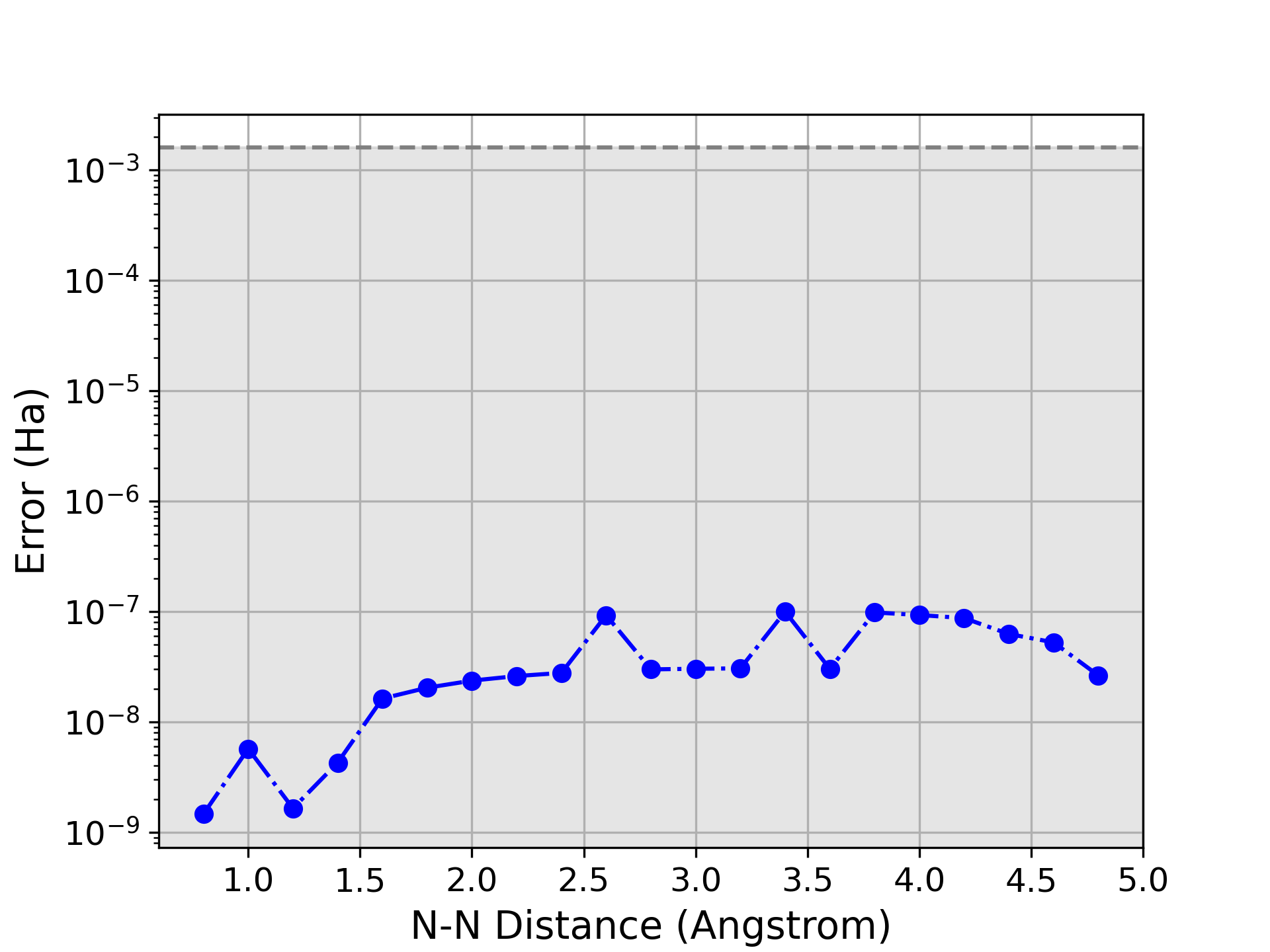}
        \caption{}
        \label{2}
    \end{subfigure}
    \caption{Comparison of ground state energy and error analysis for the $\mathrm{N_2}$ system.  (a) ADAPT-GCIM ground-state energies compared with CASCI. The dashed gray line denotes the CASCI reference. (b) Comparison of errors $E_{\mathrm{ADAPT-GCIM}} - E_{\mathrm{CASCI}}$ over entire bond distances. The shaded region below the dashed gray line indicates "chemical accuracy" as 1.59 $\times 10^{-3}$ Ha(1 kcal/mol).}
    \label{N2_combined}
\end{figure*}

\subsection{Potential energy surface of the $\mathrm{N_2}$ molecule}

We begin our benchmark suite with the potential energy surface (PES) of the $\mathrm{N_2}$ molecule, a canonical test case for assessing the ability of quantum algorithms to treat both weakly and strongly correlated electronic structure within a single system. Near its equilibrium bond length, the triple-bonded $\mathrm{N_2}$ molecule is well described by single-reference methods such as CCSD. However, as the $\mathrm{N{-}N}$ bond is stretched, multiple electronic configurations become nearly degenerate, leading to pronounced multireference character and the breakdown of single-determinant approaches. This behavior has been well documented in the literature; for example, Chan \emph{et al.}~\cite{chan2004state} showed that CCSD fails to recover accurate ground-state energies at extended bond distances, whereas DMRG remains accurate by treating all configurations on an equal footing. 
This sharp transition from weak to strong correlation, combined with the modest system size and extensive prior study, makes $\mathrm{N_2}$ an ideal first benchmark for evaluating the robustness and adaptability of ADAPT-GCIM across different correlation regimes.

Figure~\ref{N2_combined} compares the ground-state PES obtained using ADAPT-GCIM against exact CASCI results within the same active space. For each bond length, the ASF was employed to identify correlation-relevant orbitals based on single-orbital entropies. In this case, four spatial orbitals were selected from a total of 14 (obtained from an STO-6G basis), corresponding to an eight-qubit problem with 26 possible single- and double-excitation operators in the fermionic SD operator pool. Notably, ADAPT-GCIM converged using only 18 adaptively selected operators, demonstrating substantial compression relative to the full pool.

As shown in Fig.~\ref{N2_combined}(a), ADAPT-GCIM reproduces the full PES with excellent agreement to CASCI across the entire range of $\mathrm{N{-}N}$ distances, including the strongly correlated dissociation regime. The corresponding energy errors, shown in Fig.~\ref{N2_combined}(b), remain in the range of $10^{-7}$–$10^{-9}$~Ha throughout the PES, far below the threshold of chemical accuracy (1.59~$\times 10^{-3}$~Ha). Importantly, no qualitative degradation in accuracy is observed as the bond is stretched, indicating that the adaptive subspace construction successfully captures the evolving multireference character of the wavefunction.

From an algorithmic perspective, this result demonstrates that ADAPT-GCIM can seamlessly transition from single-reference to multireference regimes without requiring changes to the operator pool, reference state, or the non-optimization strategy. Although $\mathrm{N_2}$ is not itself beyond the reach of classical methods, its well-understood correlation structure provides a stringent and interpretable validation of the ADAPT-GCIM algorithmic framework. Larger active spaces for $\mathrm{N_2}$ can be constructed in a systematic manner, making this system a useful calibration point for probing algorithmic scaling and performance across increasingly complex correlation landscapes.

\subsection{Potential energy surface of the $\mathrm{FeS}$ molecule}
\begin{figure*}
    \centering
    \begin{subfigure}[b]{0.48\linewidth}
        \includegraphics[width=\linewidth]{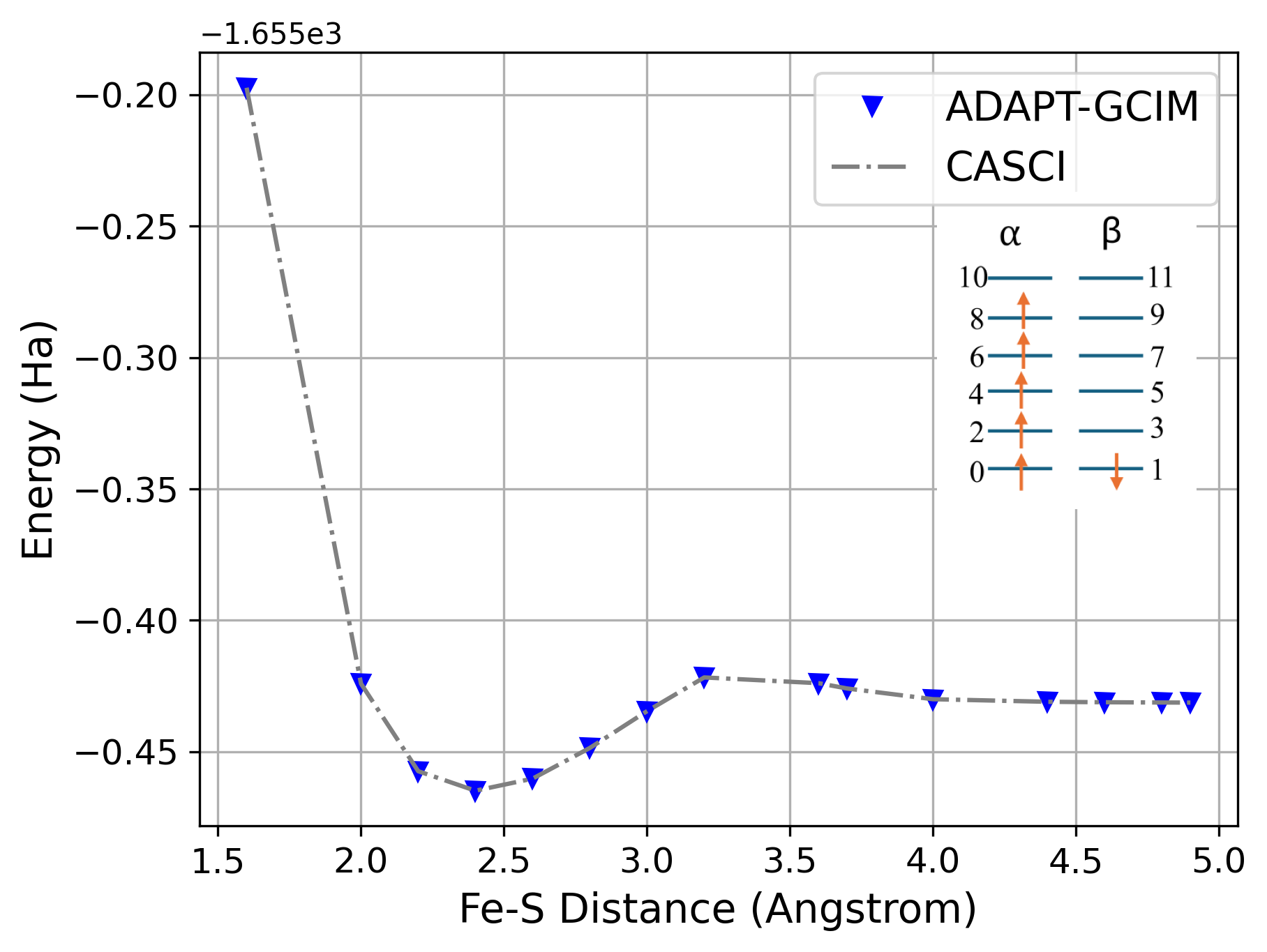}
        \caption{}
        \label{FeS_energy}
    \end{subfigure}
    \hfill
    \begin{subfigure}[b]{0.48\linewidth}
        \includegraphics[width=\linewidth]{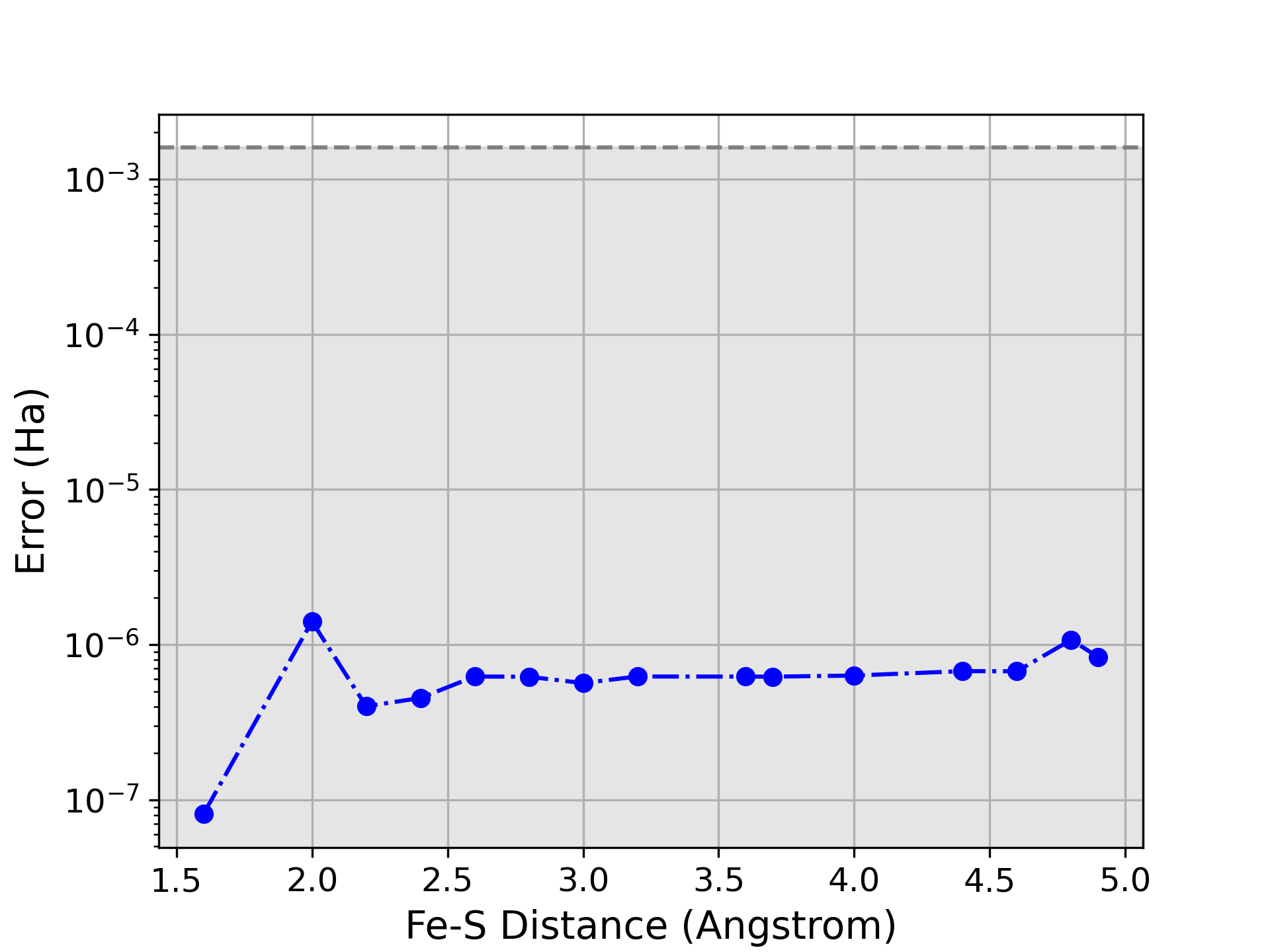}
        \caption{}
        \label{FeS_error}
    \end{subfigure}
    \caption{FeS benchmark using ADAPT-GCIM with a quintet reference state. (a) Potential energy surface comparing ADAPT-GCIM and CASCI energies within the (6e,6o) active space using the ANO-RCC-MB basis. (b) Corresponding energy errors $E_{\mathrm{ADAPT-GCIM}} - E_{\mathrm{CASCI}}$ across the Fe--S bond dissociation coordinate. The dashed grey line indicates chemical accuracy (1.59$\times10^{-3}$~Ha).}
    \label{FeS_combined}
\end{figure*}

We next consider the PES of the FeS molecule, which introduces a qualitatively different challenge from the closed-shell $\mathrm{N_2}$ system: an open-shell ground state with strong spin polarization. FeS is characterized by four unpaired electrons forming a quintet ($S=2$) ground state, making it an instructive benchmark for assessing whether adaptive quantum algorithms can accommodate high-spin reference states without manual intervention.

For this system, the ActiveSpaceFinder was used to select an active space of six electrons in six spatial orbitals, corresponding to a 12-qubit problem. Calculations were performed using the ANO-RCC-MB basis set, the minimal variant of the ANO-RCC family~\cite{Roos2004ANORCC}. This choice balances chemical realism (particularly important for transition-metal systems) with a compact orbital representation suitable for quantum simulation and benchmarking.

Unlike the closed-shell $\mathrm{N_2}$ case, FeS requires a high-spin quintet reference configuration with four unpaired electrons. This immediately exposes a fundamental limitation of conventional fermionic SD-based operator pools, which are defined relative to a closed-shell Hartree--Fock reference and rely on a well-defined occupied--virtual orbital partition. When applied directly to a high-spin open-shell reference, this construction is no longer valid, and the standard fermionic SD pool fails to provide the operator connectivity required to span the correlated ground-state manifold.
Importantly, this limitation does not arise from a violation of spin conservation, but from the incompatibility between the closed-shell fermionic SD excitation structure and the occupation pattern of an open-shell quintet reference. As a consequence, essential spin-conserving excitation pathways are absent from the primitive fermionic SD pool, preventing ADAPT-GCIM from converging to the correct ground-state energy. This behavior is therefore expected rather than pathological and underscores the necessity of constructing operator pools that are explicitly compatible with the symmetry and spin structure of the chosen reference state.
To remedy this issue, we constructed a modified operator pool by explicitly generating all symmetry-allowed single and double excitations relative to the quintet reference state. While a fully generalized singles and doubles (GSD) pool would also recover these missing operators, it comes at a significantly higher computational cost. The tailored quintet-based pool used here contains 35 operators in total and provides a minimal yet complete excitation manifold for this reference, while the GSD operator pool contains 261 operators for this space, making it a very substantial reduction. 

Figure~\ref{FeS_combined} presents a combined view of the FeS PES and the corresponding energy errors. Panel (a) compares the ground-state energies obtained using ADAPT-GCIM with exact CASCI results within the same active space, while panel (b) shows the energy differences $E_{\mathrm{ADAPT-GCIM}} - E_{\mathrm{CASCI}}$ across the full range of Fe--S bond distances. With the modified operator pool, ADAPT-GCIM converges systematically and reaches accuracies of $10^{-6}$–$10^{-7}$~Ha within 25 adaptive iterations, well below chemical accuracy.

From an algorithmic standpoint, the FeS benchmark highlights a critical but often underemphasized aspect of adaptive quantum algorithms: the choice of operator pool must be compatible with the symmetry and spin structure of the reference state. Importantly, once this compatibility is enforced, ADAPT-GCIM remains robust across the entire PES and does not require further algorithmic modification. This makes FeS a valuable intermediate benchmark that bridges closed-shell molecules and more complex open-shell transition-metal and bioinorganic systems. As with $\mathrm{N_2}$, the FeS system can be systematically extended by enlarging the active space, enabling controlled exploration of increasingly rich correlation regimes relevant to quantum utility.

\begin{table*}[t]
  \centering
  \caption{Comparison of  ADAPT-GCIM energies for $\mathrm{[2Fe\!-\!2S]}$ cluster with varying (electrons, orbitals) active space size. The singlet HF state was the input reference state. Errors in ADAPT-GCIM and ADAPT-VQE are with CASCI energies as reference. Li \emph{et al.}~\cite{doi:10.1021/acs.jctc.7b00270} reported DMRG results for the cluster at (30e,20o) to be -116.6056091 Ha}
  \begin{tabular}{lllll}
    \hline
    System (\textit{electrons, orbitals}) & CASCI Energy(\textit{Ha}) & Error in ADAPT-GCIM (\textit{Ha}) & Error in ADAPT-VQE(\textit{Ha}) \\
    \hline
    $[$2Fe-2S$]$ (4e,4o)       & -116.154749    & 1.538$\times 10^{-8}$    & 2.106$\times 10^{-7}$ (14 iterations)\\
    $[$2Fe-2S$]$ (8e,6o)       & -116.158572    & 2.764$\times 10^{-4}$ & 1.089$\times 10^{-4}$  (42 iterations)\\
    $[$2Fe-2S$]$ (6e,6o)       & -116.156259    & 3.357$\times 10^{-5}$ & 1.252$\times 10^{-5}$  (54 iterations)\\
    \hline
  \end{tabular}
  \label{5}
\end{table*}

\subsection{Ground-state energy of the $\mathrm{[2Fe\!-\!2S]}$ cluster}

We next turn to the $\mathrm{[2Fe\!-\!2S]}$ cluster, an open-shell transition-metal system that introduces a qualitatively richer correlation structure than the diatomic benchmarks discussed above. Iron–sulfur clusters play a central role in biological electron-transfer chains and catalytic active sites, and their electronic structure is dominated by strong correlation, spin coupling, and near-degeneracy effects that go well beyond toy-model chemistry.
Recent work by M{\"o}rchen \emph{et al.}~\cite{morchen2024classification} introduced a classification scheme for iron–sulfur clusters based on orbital entanglement measures, demonstrating that despite their strong correlation, certain clusters admit substantial compression into compact active spaces. Among these, the $\mathrm{[2Fe\!-\!2S]}$ cluster emerges as a particularly attractive candidate: it retains essential multireference character while remaining sufficiently compact to enable controlled benchmarking. Complementary studies by Lee \emph{et al.}~\cite{Lee2023Evaluating} further showed that, among a hierarchy of iron–sulfur clusters ranging from $\mathrm{[2Fe\!-\!2S]}$ to $\mathrm{[4Fe\!-\!4S]}$ and ultimately FeMoco, the $\mathrm{[2Fe\!-\!2S]}$ cluster exhibits the largest overlap between the best single determinant and the exact ground state. This overlap decreases rapidly for larger clusters, reflecting a transition to increasingly delocalized multireference wavefunctions with no dominant determinant.
As a result, the $\mathrm{[2Fe\!-\!2S]}$ cluster occupies a crucial intermediate regime: it is chemically realistic and strongly correlated, yet still amenable to classical methods such as DMRG~\cite{doi:10.1021/acs.jctc.7b00270}, making it an ideal stepping stone for benchmarking quantum algorithms intended to address larger bioinorganic systems, including FeMoco.

In this work, ground-state energies were computed using the one- and two-electron integrals provided by Li \emph{et al.}~\cite{doi:10.1021/acs.jctc.7b00270}. The full $\mathrm{[2Fe\!-\!2S]}$ cluster contains 120 electrons, and the reference benchmark reported in Ref.~\cite{doi:10.1021/acs.jctc.7b00270} employs an active space of 30 electrons in 20 spatial orbitals (30e,20o), corresponding to a 40-qubit Hamiltonian. While this large active space is beyond the scope of classical quantum-circuit simulation, it provides a chemically well-motivated target that contextualizes reduced active-space calculations.
Rather than attempting a variational convergence towards the (30e,20o) benchmark, we use a series of reduced active spaces as feasibility-scale surrogates to assess the performance of ADAPT-GCIM on a realistic iron–sulfur cluster Hamiltonian. Specifically, we carried out simulations for the (4e,4o), (6e,6o), and (8e,6o) active spaces, with the resulting ground-state energies summarized in Table~\ref{5}. These reduced models are not nested within a single orbital hierarchy and therefore should not be interpreted as a systematically convergent sequence towards the large active-space result.
For the smallest active space, (4e,4o), most correlation-relevant orbitals are effectively frozen, limiting the amount of static and dynamic correlation that can be recovered and yielding a comparatively high total energy. Expanding the active space to (6e,6o) introduces additional near-degenerate orbitals and increases variational flexibility, resulting in a slightly lower energy.
The (8e,6o) case can also be seen to result in slightly reduced energy, which aligns with the general expectation.
%The (8e,6o) case, while including more electrons within the same orbital manifold, represents a distinct effective Hamiltonian rather than a strict extension of the (6e,6o) model. As a consequence, its absolute energy need not be lower, reflecting changes in electron–electron repulsion and correlation balance rather than a failure of the algorithm.

Overall, these results demonstrate that ADAPT-GCIM can be applied in a controlled and stable manner to progressively more complex active-space models of a strongly correlated transition-metal cluster. Although the reduced active spaces do not quantitatively reproduce the large-active-space DMRG energy, they capture increasingly rich correlation physics while remaining tractable for quantum simulation. Importantly, this progression is achieved without modifying the underlying algorithmic framework, highlighting the extensibility of ADAPT-GCIM across different effective Hamiltonians.

From a broader perspective, the $\mathrm{[2Fe\!-\!2S]}$ benchmark serves as a proof-of-principle that ADAPT-GCIM can handle chemically realistic, multi-center correlated systems beyond diatomic or single-metal examples. In this sense, it provides a critical methodological bridge between small-molecule benchmarks and more challenging bioinorganic targets, such as larger iron–sulfur clusters and ultimately FeMoco.

\subsection{Potential energy surface of $\mathbf{U_2}$}

\begin{figure*}
    \centering
    \begin{subfigure}[b]{0.48\linewidth}
        \centering
        \includegraphics[width=\linewidth]{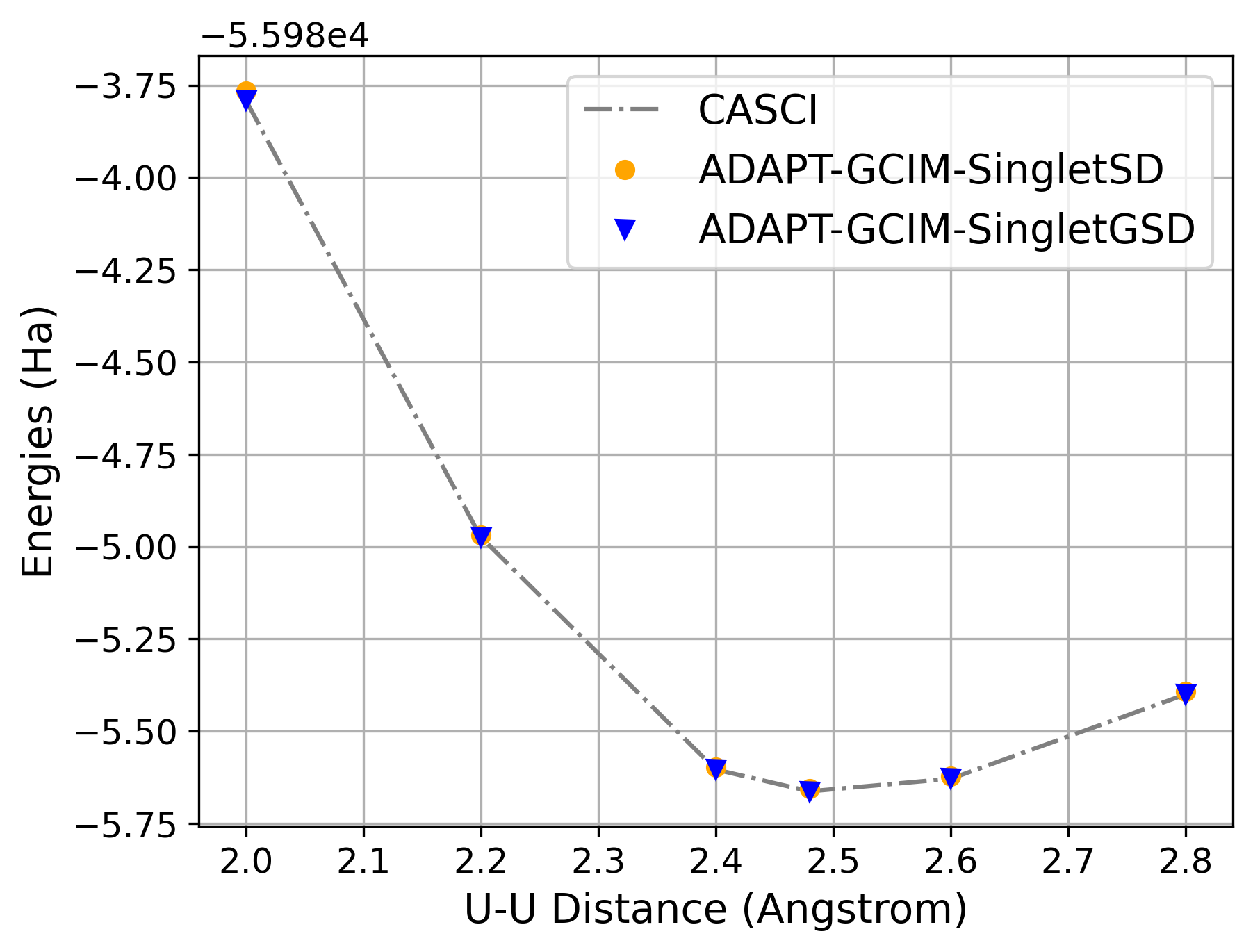}
        \caption{}
        \label{fig:U2_energy}
    \end{subfigure}
    \hfill
    \begin{subfigure}[b]{0.48\linewidth}
        \centering
        \includegraphics[width=\linewidth]{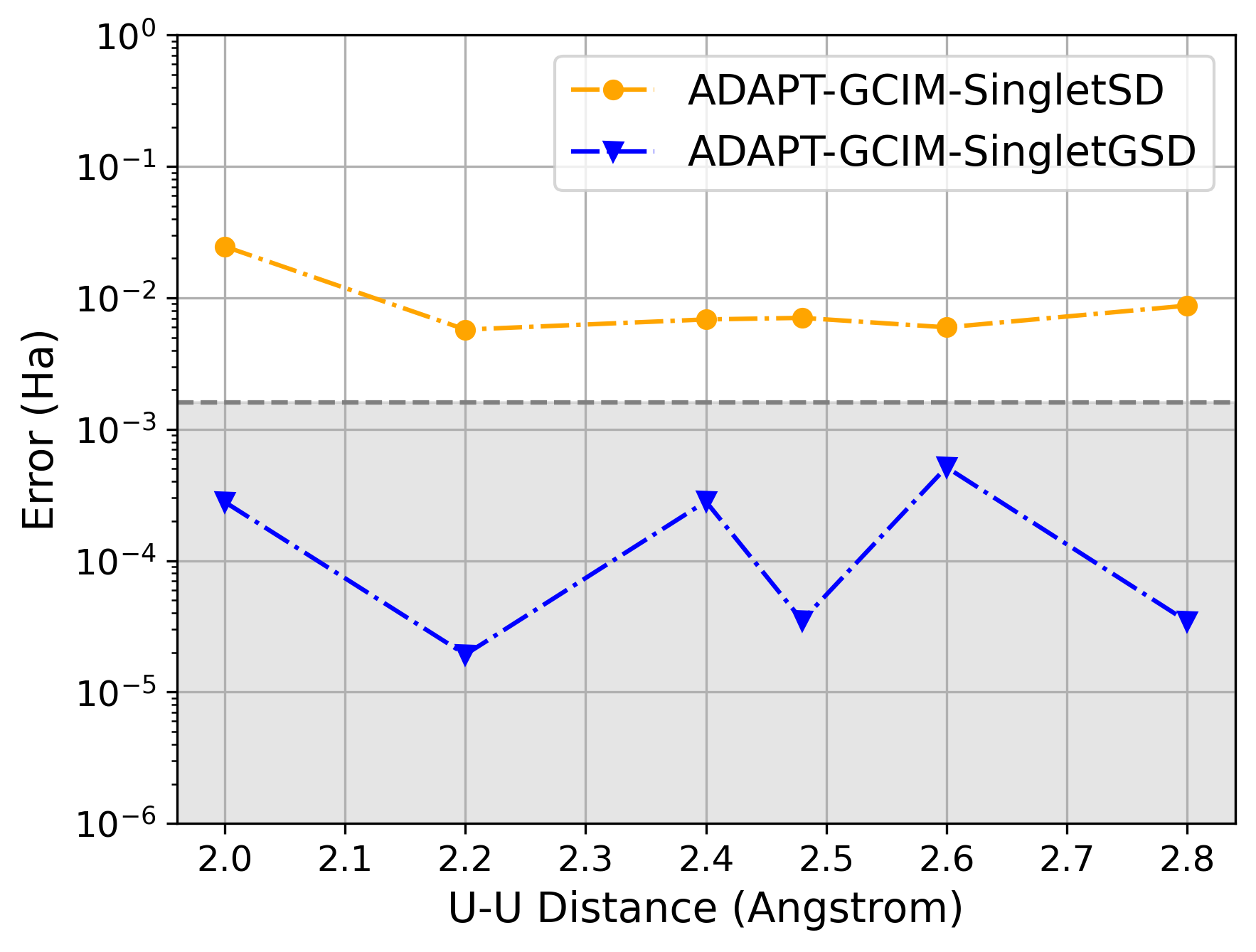}
        \caption{}
        \label{fig:U2_error}
    \end{subfigure}
    \caption{
    Potential energy surface and accuracy analysis for $\mathrm{U_2}$ in the (6e,6o) active space with a spin-free X2C-1e Hamiltonian and ANO-RCC-MB basis.
    (a) Ground-state potential energy surface obtained from ADAPT-GCIM using SingletSD and SingletGSD operator pools, compared against CASCI reference energies.
    (b) Corresponding energy errors $E_{\mathrm{ADAPT\text{-}GCIM}} - E_{\mathrm{CASCI}}$ along the dissociation coordinate.
    The dashed grey line indicates chemical accuracy (1.59 $\times 10^{-3}$ Ha).
    SingletGSD achieves substantially improved accuracy by enhancing connectivity within the singlet manifold.
    }
    \label{fig:U2_combined_energy_error}
\end{figure*}

\subsubsection{Background and methodological scope}

The $\mathrm{U_2}$ molecule represents one of the most challenging diatomic systems in quantum chemistry, characterized by strong multireference character, near-degeneracy effects, and a delicate interplay between static correlation, relativistic effects, and spin coupling. Actinide--actinide bonding in $\mathrm{U_2}$ involves partially occupied $5f$ and $6d$ orbitals, leading to a highly entangled electronic structure that is extremely sensitive to the choice of active space and the treatment of relativistic and dynamical correlation effects. 

Previous high-level studies have reported conflicting pictures of the $\mathrm{U_2}$ bond. Gagliardi \emph{et al.}~\cite{gagliardi2005quantum} employed a relativistic CASSCF(6e,20o) treatment followed by CASPT2 with spin--orbit coupling included perturbatively and reported a quintuple bond. In contrast, Knecht \emph{et al.}~\cite{knecht2019relativistic} treated spin--orbit coupling variationally using a two-component X2C Hamiltonian within a RASSCF framework and concluded a lower bond order. These discrepancies underscore the extreme sensitivity of $\mathrm{U_2}$ to methodological choices and motivate the development of systematically improvable many-body approaches. There are only a few preliminary studies on the use of quantum computers in relativistic quantum chemistry, with the first proposal in Refs.~\cite{veis2012relativistic}. Other numerical studies include the study of atomic fine structure splitting~\cite{sugisaki2023bayesian} and molecular electric dipole moments~\cite{chawla2025relativistic}. Perhaps the only other study that we could find that proposes the study of actinide chemistry on quantum computers is recent in Ref. \cite{sorathia2025quantum}, but the actinide-actinide bonds and other challenging problems in heavy-element chemistry that could be critical for quantum utility have so far been unexplored. These problems are challenging both because the actinide-actinide bonds provide challenging strong correlation problems in large active spaces, and rigorous treatment of electron correlation by including spin-orbit coupling and resulting spin-symmetry breaking results in doubling of the size of the active space (if electron correlation is included at the spinor level). Since quantum computers already map spin orbitals to qubits (in Jordan-Wigner mapping), they could be ideal testbeds for rigorous treatment of electron correlation with relativistic effects in challenging heavy element chemistry problems.

The goal of the present study is not to resolve the quantitative bonding debate in $\mathrm{U_2}$, which would require much larger active spaces and fully relativistic treatments. Rather, we use $\mathrm{U_2}$ as a stringent algorithmic stress test for ADAPT-GCIM, focusing on how algorithm design impacts the representation of a strongly multireference singlet ground state.

\subsubsection{Computational setup and reference CASCI surface}

To incorporate scalar relativistic effects, all electronic-structure calculations were carried out using the spin-free one-electron exact two-component  (SF-X2C-1e) Hamiltonian~\cite{cheng2014perturbative,liu2021relativistic,lu2022exact,knecht2022exact,Asthana2019,10.1063/1.3159445} with the ANO-RCC-MB basis set. In this formalism, the four-component Dirac Hamiltonian is rigorously decoupled into an effective two-component representation at the one-electron level, providing an exact treatment of scalar relativistic effects while substantially reducing the computational cost relative to a full four-component approach~\cite{iliavs2007infinite,saue2011relativistic}. Although spin–orbit coupling is known to be important for actinide systems such as U$_2$, it is neglected here for computational simplicity, and the present work therefore focuses on scalar relativistic contributions to the electronic structure. This is an efficient and internally consistent framework for incorporating leading relativistic effects in combination with the correlated electronic-structure methods employed in this study.
A reduced active space of six electrons in six spatial orbitals (6e,6o) was employed to keep the problem tractable for quantum simulation while retaining strong static correlation.

A restricted open-shell Hartree--Fock (ROHF) solution was used as the starting point, followed by a CASSCF calculation at a reference geometry of 2.5~\AA. The resulting active orbitals were projected onto other geometries along the dissociation coordinate, and CASCI calculations were performed with the total spin constrained to be singlet. This procedure yields a smooth singlet potential energy surface with an equilibrium bond distance in the range of 2.5--2.6~\AA, qualitatively consistent with more sophisticated relativistic treatments~\cite{knecht2019relativistic}. We note that the ActiveSpaceFinder tool could not be employed in this case because of the challenging nature of orbitals involved in the active space, and we selected the active space after a single CASSCF calculation as discussed above. A comparison of ADAPT-GCIM potential energy surfaces obtained using different operator pools against this CASCI reference is shown in Fig.~\ref{fig:U2_combined_energy_error}.

\begin{figure}[t]
    \centering
    \includegraphics[width=0.9\linewidth]{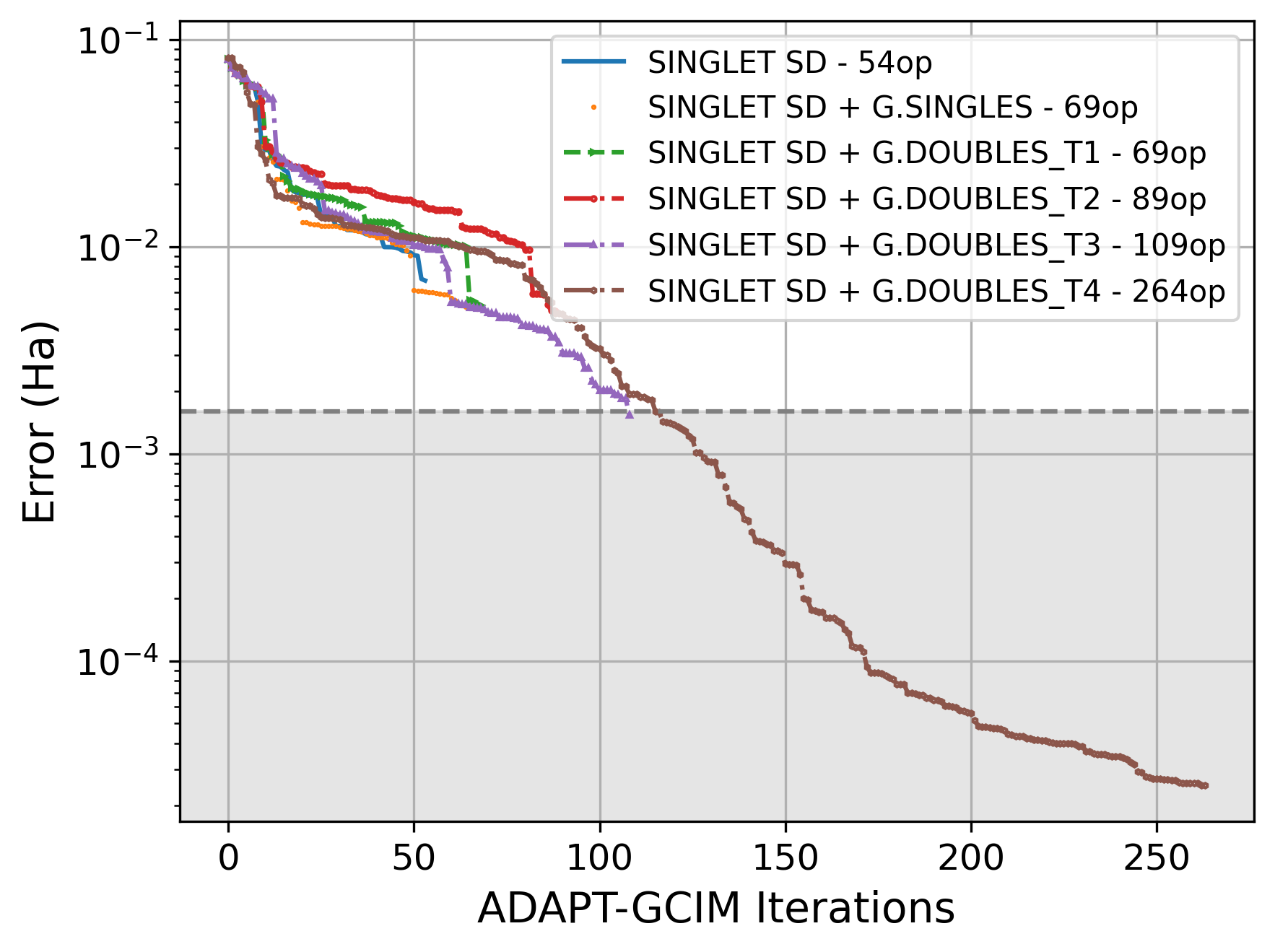}
    \caption{
    Comparison of ADAPT-GCIM convergence for $\mathrm{U_2}$ using SingletSD alone and SingletSD augmented with specific classes of generalized doubles.
    Inclusion of T4-type generalized singles and doubles, which directly couple electrons between different spatial orbitals, enables convergence to chemical accuracy with significantly fewer operators.
    This highlights the critical role of operator-pool design in efficiently capturing strong multireference correlation.
    Each line in the plot stops when the operators in the pool are exhausted.
    }
    \label{fig:U2_GD_types}
\end{figure}

\subsubsection{Failure of SingletSD pool and existence of strong higher-rank correlations}

We first applied ADAPT-GCIM using a singlet-adapted singles-and-doubles (SingletSD) operator pool. Despite explicit spin adaptation, this operator pool fails to achieve chemical accuracy across the $\mathrm{U_2}$ potential energy surface (see Fig. \ref{fig:U2_combined_energy_error}). This is the first system in our study which has demonstrated a failure of the SingletSD pool in reaching chemical accuracy for ADAPT-GCIM, indicating the difficulty of recovering electron correlation in the (6e,6o) active space of U$_2$ molecule. We also see in the same figure that the SingletGSD pool is able to solve the active space problem to sufficient accuracy. We carry out a careful analysis below to investigate this phenomenon.

This failure does not arise from missing spin-flip processes or spin contamination; instead, it can be traced to the compactness of the ADAPT-GCIM ansatz. The subspace created by ADAPT-GCIM introduces higher-rank excited configurations, but there is not enough variational flexibility to reach the higher-rank excited configurations in the final state.

\begin{figure}[t]
    \centering
    \includegraphics[width=\linewidth]{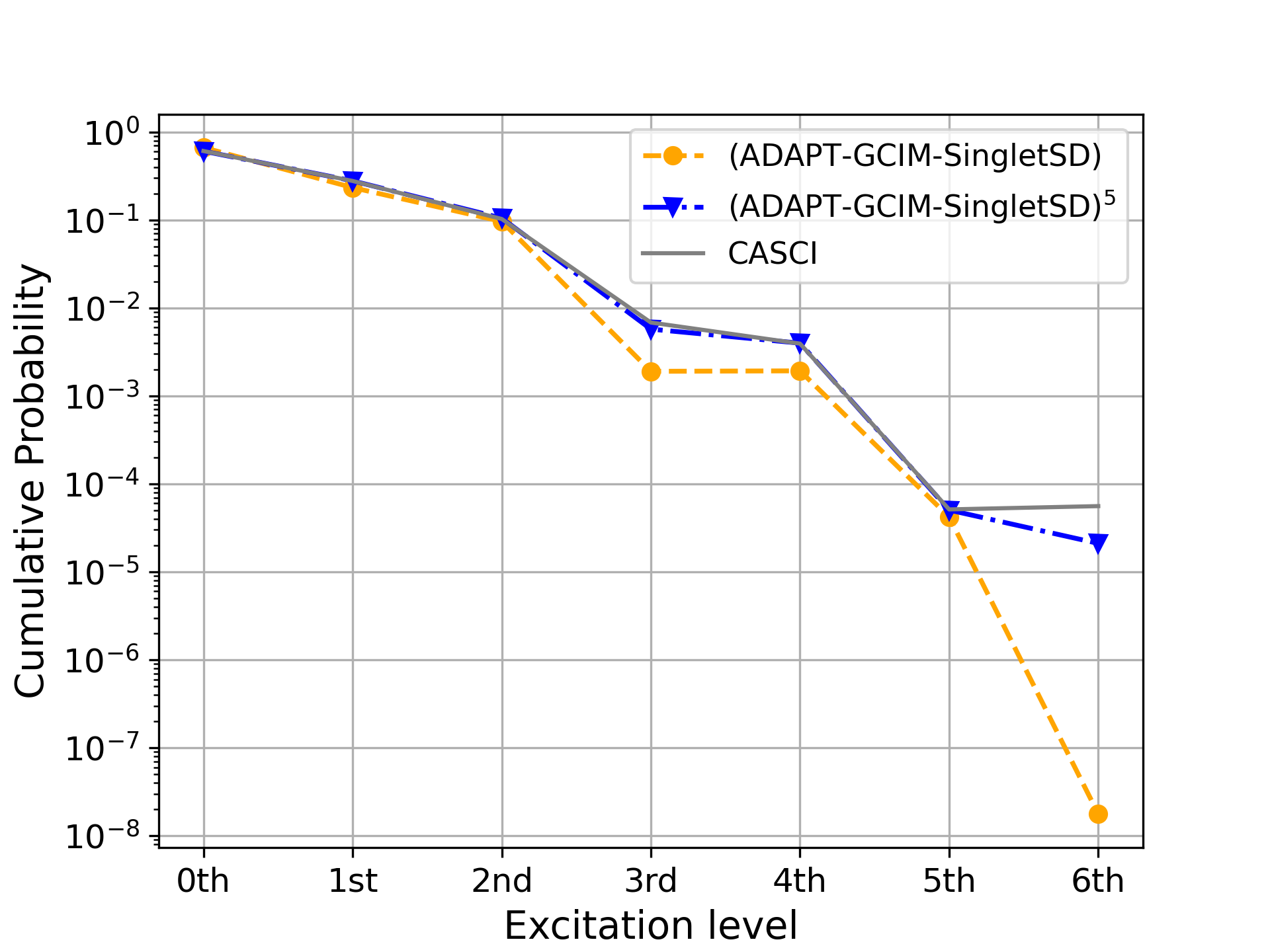}
    \caption{
    Total probability of n-body excited determinants, comparing CASCI, (ADAPT‑GCIM SingletSD) and (ADAPT-GCIM-SingletSD)$^5$ wavefunctions as a function of excitation rank. Determinants are grouped by excitation rank relative to the reference Hartree-Fock configuration, and the total probability is accumulated for each excitation order.}
    \label{fig:U2_wavefncomp}
\end{figure}

\begin{figure}[t]
    \centering
    \includegraphics[width=\linewidth]{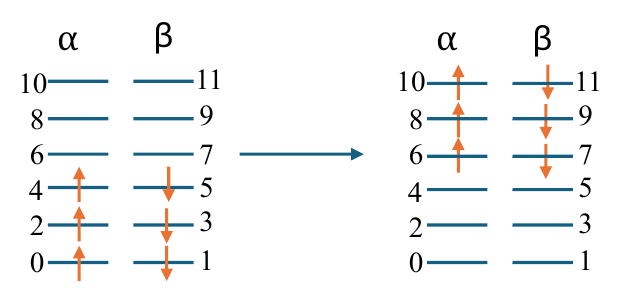}
    \caption{Example of a high-order excitation required in the case of U$_2$(6e,6o) molecule. It has a significant contribution of the sixth-order excited configuration from the Hartree-Fock reference in the exact ground state, with the energy contribution of more than 1 Ha.} 
    \label{u2excitation}
\end{figure}

In Fig. \ref{fig:U2_wavefncomp}, we present the accumulated probability of different orders of excited configurations in exact CASCI, ADAPT-GCIM-SingletSD and (ADAPT-GCIM-SingletSD)$^5$.
A detailed list of probabilities and energy contributions from major configurations of excitation orders one to six are provided in the SI.
We note that ADAPT-GCIM-SingletSD fails to reach close accuracies in accumulated probabilities in third to sixth order excitations.
Looking at the energy contributions in SI, it is clear that even the sixth-order excited configuration (see Fig. \ref{u2excitation}) is essential to be resolved accurately to reach chemical accuracy. One cannot reach chemical accuracy without achieving reasonable accuracy in these configurations and higher order.

Although ADAPT-GCIM introduces these higher-rank excitations through its operator product expressions in Eq. \eqref{eq:gcimmain}, it requires enough parameters to reach these states from the single and double type excitations used in the generation of the subspace. If there are insufficient operators added, it will result in a `parameter starvation' and result in a lack of variational flexibility to reach the higher-rank excited configurations. 

This problem can be solved by using the SingletGSD operator pool in Fig. \ref{fig:U2_combined_energy_error}.
This pool will have generalized single and double operators with a total of 330 operators compared with 54 of the singletSD pool. Although SingletGSD can reach beyond chemical accuracy in this case, the increase in the number of operators in a pool is unfavorable because it directly increases the number of shots in the algorithm by O(N$^6$). With a 6-fold increase in the number of operators in the pool, the shot count cost increases $\sim 5\times10^4$ times, which is a very large increase.
Due to this, a compact pool, like singletSD, is preferred over a larger pool of operators. 

  In Fig. \ref{fig:U2_GD_types}, we investigate if there is a key operator type in the SingletGSD pool that is capable of capturing the missing physics of U$_2$. We plot the convergence of  (i) SingletSD pool, (ii) SingletSD pool + singlet generalized singles, (iii) SingletSD pool + singlet generalized doubles of type T$_1$, (iv) SingletSD pool + singlet generalized doubles of type T$_2$, (v) SingletSD pool + singlet generalized doubles of type T$_3$, (vi) SingletSD pool + singlet generalized doubles of type T$_4$. These pools are designed by adding each part of the SingletGSD operator pool individually on top of the singletSD pool, such that if there is any single type of missing operators, that would have a dramatic positive impact in that modified pool's performance.
It can quickly be observed that the performance of all the above pools is similar in terms of ``the average correlation captured per operator''. Each type of operator achieves an energy that is almost directly proportional to the number of operators. This shows that there is little benefit in adding any specific type of operator in the operator pool; instead, the number of operators and the resulting increase in the number of parameters (variational flexibility) are the more important aspects.

\begin{figure}[t]
    \centering
    \includegraphics[width=\linewidth]{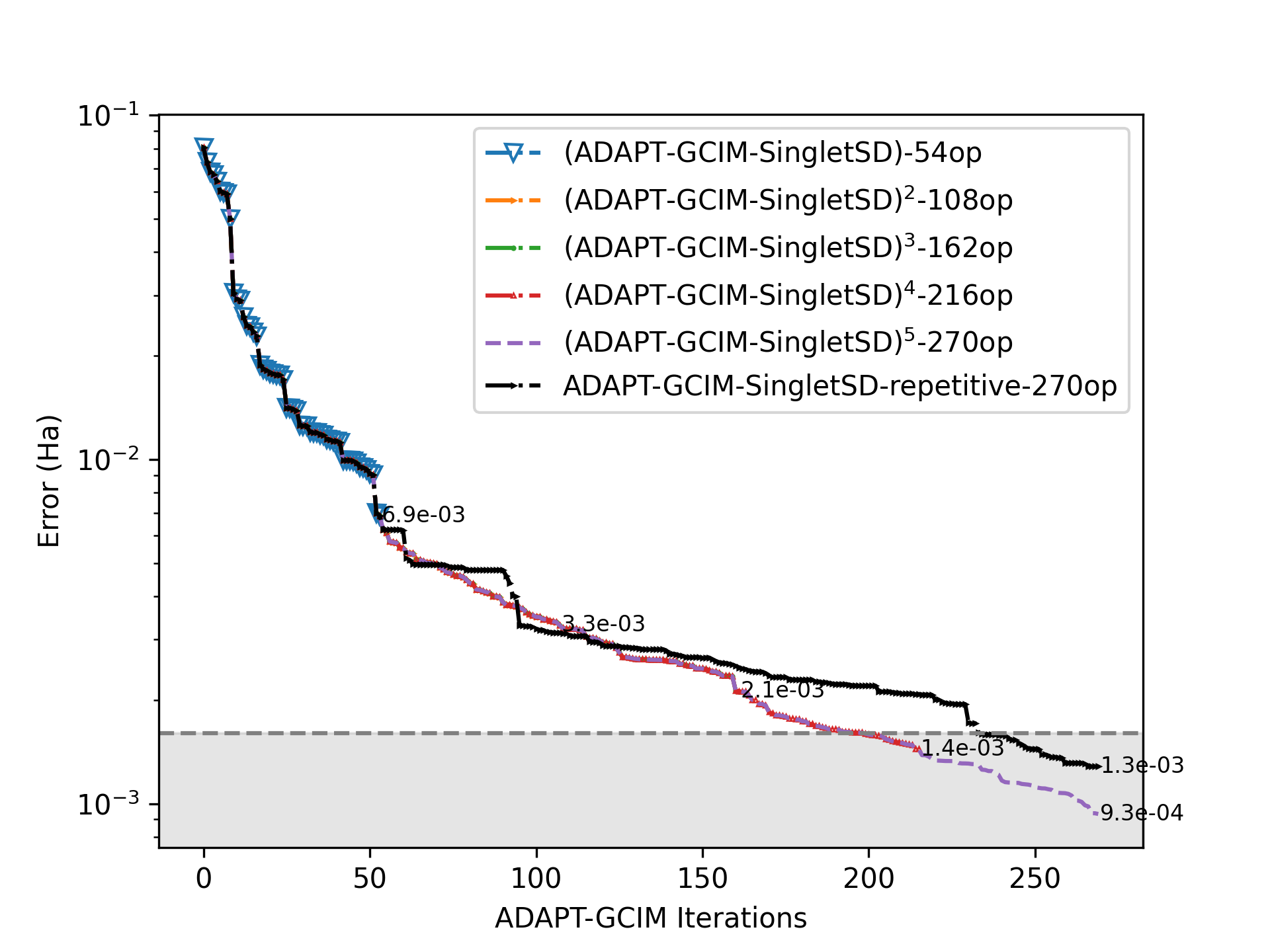}
    \caption{Comparison of errors $E_{\mathrm{ADAPT-GCIM}} - E_{\mathrm{CASCI}}$ for $\mathrm{U_2}$ system with SingletSD operator pools. Operators are repeated in ADAPT-GCIM calculations once the pool is drained completely.  The black dashed lines refer to ADAPT-GCIM results obtained from the repetition of operators after the completion of a standard ADAPT-GCIM calculation. The dashed grey line and the section below refer to "chemical accuracy" as 1.59 $\times 10^{-3}$ Ha(1 kcal/mol) }
    \label{fig:U2repeat}
\end{figure}

\begin{figure*}[t]
    \centering
    \begin{subfigure}[b]{0.48\linewidth}
        \centering
        \includegraphics[width=\linewidth]{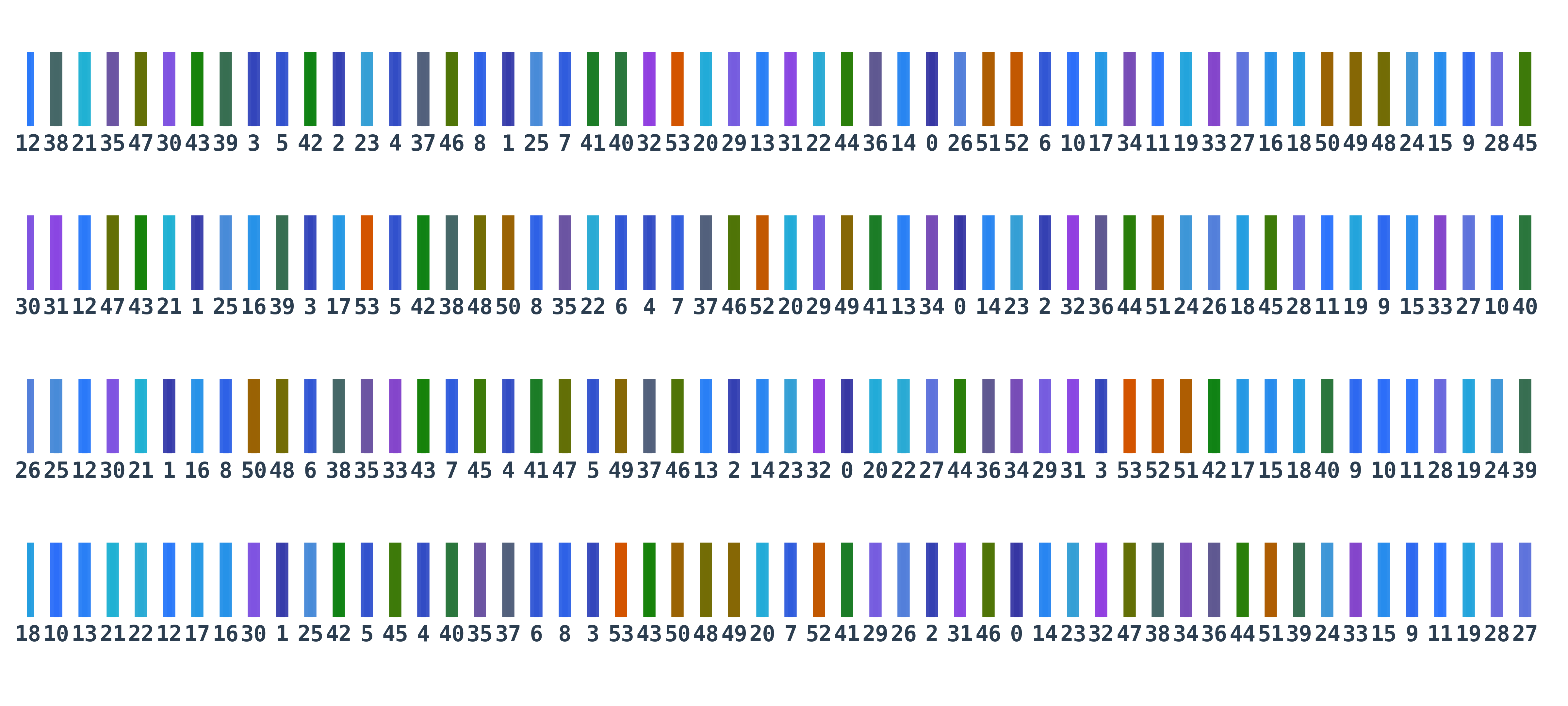}
        \caption{ADAPT-GCIM$^n$}
        \label{fig:U2_generic_SD5}
    \end{subfigure}
    \hfill
    \begin{subfigure}[b]{0.48\linewidth}
        \centering
        \includegraphics[width=\linewidth]{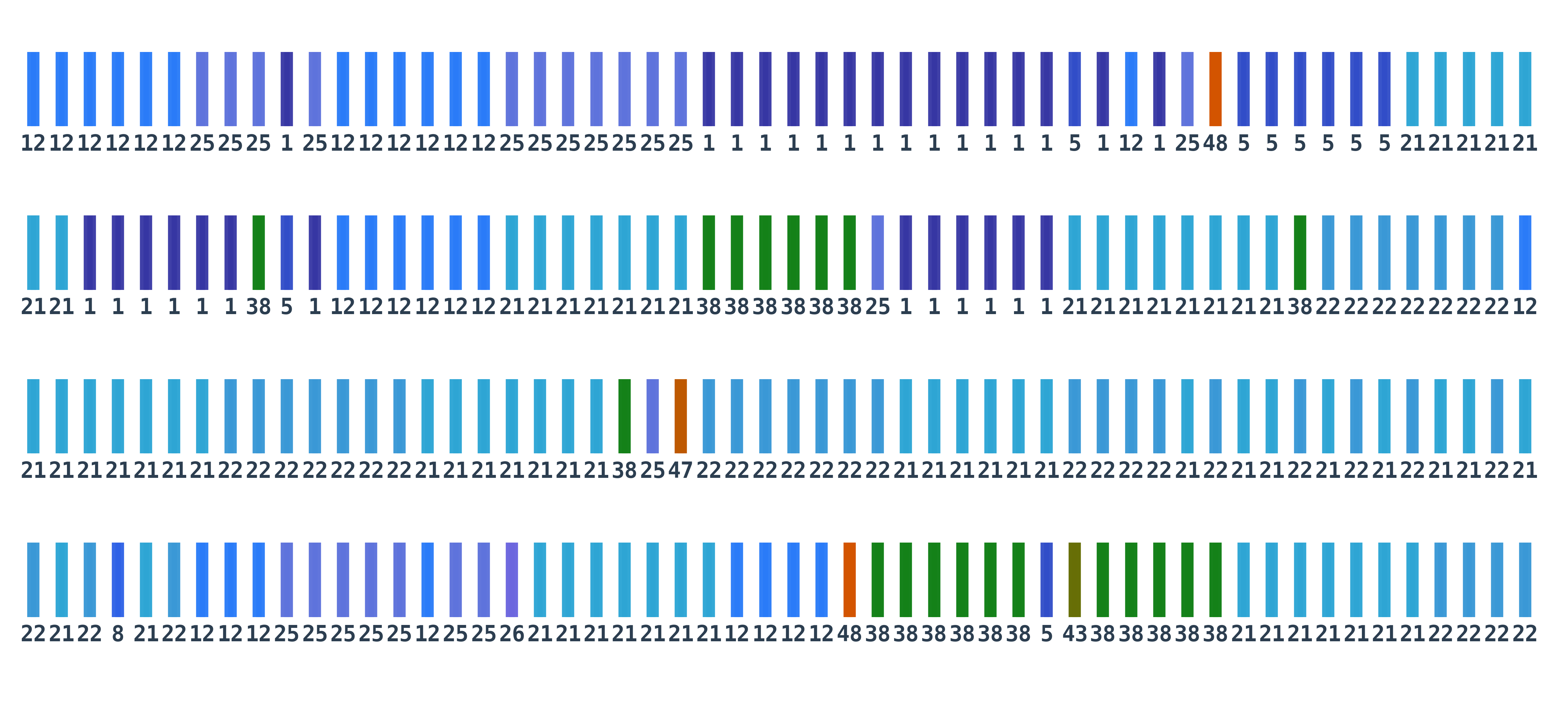}
        \caption{ADAPT-GCIM-repetitive}
        \label{fig:U2_generic_SDrep}
    \end{subfigure}
    \caption{
    Operator addition order in ansatz for (a) (ADAPT-GCIM)$^5$ and (b) (ADAPT-GCIM-repetitive). The first 54 operators are the same for both ansatz, hence they have been removed for more clarity. The operator pool is refreshed for every ADAPT-GCIM cycle for the ansatz in (a). In the ansatz (b), the operators are allowed to repeat within each cycle. 
    }
    \label{fig:U2_generic_comb}
\end{figure*}

\subsubsection{Operator repetition as a workaround}

To capture the missing physics of the system, we designed two efficient strategies in ADAPT-GCIM: (i) repeating ADAPT-GCIM ansatz n times (referred to as (ADAPT-GCIM)$^n$ and (ii) allowing repetition of SingletSD operators in the ADAPT-GCIM algorithm after completion of a single complete run of ADAPT-GCIM (referred to as ADAPT-GCIM repetitive). 
 As a minor technical point, we note that the two strategies can have a convergence-based stopping criterion, such that it can be stopped once the difference between two iterations reaches a threshold. For analysis, no such threshold is chosen in this study.

As noted by Evangelista \emph{et al.}~\cite{evangelista2019exact,evangelista2018perspective}, products of repeating single- and double-excitation operators can, in principle, generate effective higher-rank excitations and thereby span a more complete configuration space. Consistent with this expectation, repeating ADAPT-GCIM cycles using the SingletSD operator pool systematically lowers the energy by implicitly constructing higher-body excitations through operator products.
Fig. \ref{fig:U2repeat} shows the performance of the two strategies. Both strategies reach chemical accuracy and show promising convergence characteristics. We note that the overlap improves in each iteration, and the ADAPT-GCIM converges to the exact CASCI state iteratively after each iteration. The table of higher-rank excitations in SI also confirms this fact. It can be seen that the amplitudes of the higher-rank excitations reach closer to the amplitudes in the exact CASCI solution after (ADAPT-GCIM)$^4$ (216 operators) and around 250$^{\text{th}}$ iteration in repeating operator ADAPT-GCIM. 

Fig.\ref{fig:U2_generic_comb} represents the operator ordering in (ADAPT-GCIM-SingletSD)$^5$ and ADAPT-GCIM-SingletSD-repetitive. The image shows clear differences in the kind of operators added in the ans\"{a}tze while still achieving similar results, demonstrating the importance of parameter addition. The (ADAPT-GCIM-SingletSD)$^5$ adds unique operators once in each cycle and shows a diverse operator addition pattern, while the ADAPT-GCIM-SingletSD-repetitive strategy repeatedly adds the same operator many times, shown in the figure through many adjacent operator lines of the same color.

These two repeating operator strategies require a larger number of iterations (and resulting operator additions) in ADAPT-GCIM to converge, but require significantly reduced quantum computer time and measurements compared with the SingletGSD pool. This is because the addition of operators require an increase of shot count, which scales as O(N$^4$i), where `N' is the number of qubits and `i' is the number of iterations, while increasing operators in the operator pool increases shot counts with an O(N$^6$i). Due to these factors, these two strategies may be more favourable than a larger pool like SingletGSD. However, the two ADAPT-GCIM repeating operator strategies are slower in convergence than the use of the SingletGSD pool, and may have a higher chance of stalling, which is seen in ADAPT-VQE in Ref. \cite{grimsley2023adaptive}.

\section{Conclusion}\label{conclusion}

In this work, we introduced a curated hierarchy of chemically meaningful benchmark systems for evaluating quantum algorithms in regimes of electronic correlation that extend well beyond minimal toy models. By spanning multireference bond breaking (N$_2$), high-spin transition-metal chemistry (FeS), bioinorganic metalloclusters ([2Fe--2S]), and heavy-element actinide bonding (U$_2$), this benchmark set provides a systematic progression across qualitatively distinct correlation landscapes that are more representative of the molecular problems quantum computing algorithms are ultimately intended to address.

As a concrete realization, we developed and applied a practical black-box workflow that integrates automated active-space selection using the entropy-based ActiveSpaceFinder approach and enables systematic benchmarking of automated and adaptive quantum algorithms. Using ADAPT-GCIM as the primary framework, we demonstrated that compact generator-coordinate--inspired subspace expansions can achieve high accuracy for several challenging correlation regimes within chemically motivated active spaces. Equally importantly, the benchmark suite reveals general design constraints and failure modes that emerge in strongly correlated chemistry, including limitations associated with generic operator pools, the need for problem-adapted operator sets in high-spin manifolds, and the increasing importance of high-rank excitations in heavy-element bonding even within reduced valence active spaces. 
We emphasize that some failure modes—such as parameter starvation and operator-pool incompleteness—are driven by the underlying chemical correlation structure rather than by any algorithm-specific deficiency, and are therefore expected to arise broadly across variational quantum-algorithm frameworks.

Beyond the performance of any single method, the central outcome of this study is a chemically grounded benchmarking framework that enables meaningful, cross-system assessment of quantum algorithms in regimes where the structure of electronic correlation qualitatively changes. To support systematic benchmarking and reproducible cross-method comparisons, the second-quantized Hamiltonians for all systems studied in this work are made openly available.
Beyond their immediate benchmarking value, the problems introduced here are designed to prepare quantum algorithms for systematic scaling towards regimes that are inaccessible to classical computational chemistry. The [2Fe–2S] cluster serves as a natural precursor to larger and more complex iron–sulfur systems, ultimately targeting biologically and catalytically important species such as FeMoco. In parallel, the exploration of different active-space choices for U$_2$ provides a pathway towards addressing some of the most challenging problems in heavy-element quantum chemistry, including actinide–actinide bonding, whose rigorous treatment lies beyond the reach of current classical methods.
To our knowledge, this work represents the first systematic application of quantum algorithms to actinide–actinide bonded systems, with U$_2$ serving as a concrete and demanding example. In particular, the six-orbital active-space treatment of U$_2$ constitutes an exceptionally challenging benchmark due to the significant weight of high-rank excited configurations in the exact wavefunction, making it a stringent test for any variational quantum algorithm.

These results emphasize that progress towards quantum utility in chemistry will require not only improvements in hardware and algorithmic efficiency, but also chemically informed strategies that explicitly account for the correlation regimes more representative of the target problem. 
The framework introduced here provides a scalable basis for a structured, open database of quantum chemistry benchmark problems with standardized, easy-to-use Hamiltonians. Our goal is for this resource to support the systematic evaluation of present and future quantum algorithms, accelerate methodological development, and provide a clear pathway towards achieving quantum utility in the quantum chemistry domain. 
Looking forward, the benchmark hierarchy and workflow presented here provide a practical foundation for future studies that incorporate noise, finite sampling, and hardware constraints, as well as extensions to excited-state properties and larger active spaces.

\section{Data availability}
The Hamiltonians in Jordan Wigner representation for the benchmarking systems and the code for generating and using them are made available open source in the GitHub repository at: https://github.com/srivathsanps-quantum/Benchmark-QC 

\section{Acknowledgements}
AA, SPS acknowledge NSF, award number 2429752, and  NSF, award number 2427046, for support. BP and VA are supported by a U.S. Department of Energy, Office of Science, Early Career Research Program award in the Basic Energy Sciences, Division of Chemical Sciences, Geosciences, and Biosciences, Computational and Theoretical Chemistry program under FWP 83466. Pacific Northwest National Laboratory (PNNL) is operated by Battelle for the U.S. DOE under contract DE-AC05-76RL01830. All authors acknowledge the UND Computational Research Center for computing resources. SPS would like to thank Sudharsan for the useful discussions.

\bibliographystyle{achemso}  % ACS/JACS style
\bibliography{main}% Produces the bibliography via BibTeX.

\section{Representative image}
\begin{figure*}
    %\centering
    \includegraphics[width=\linewidth]{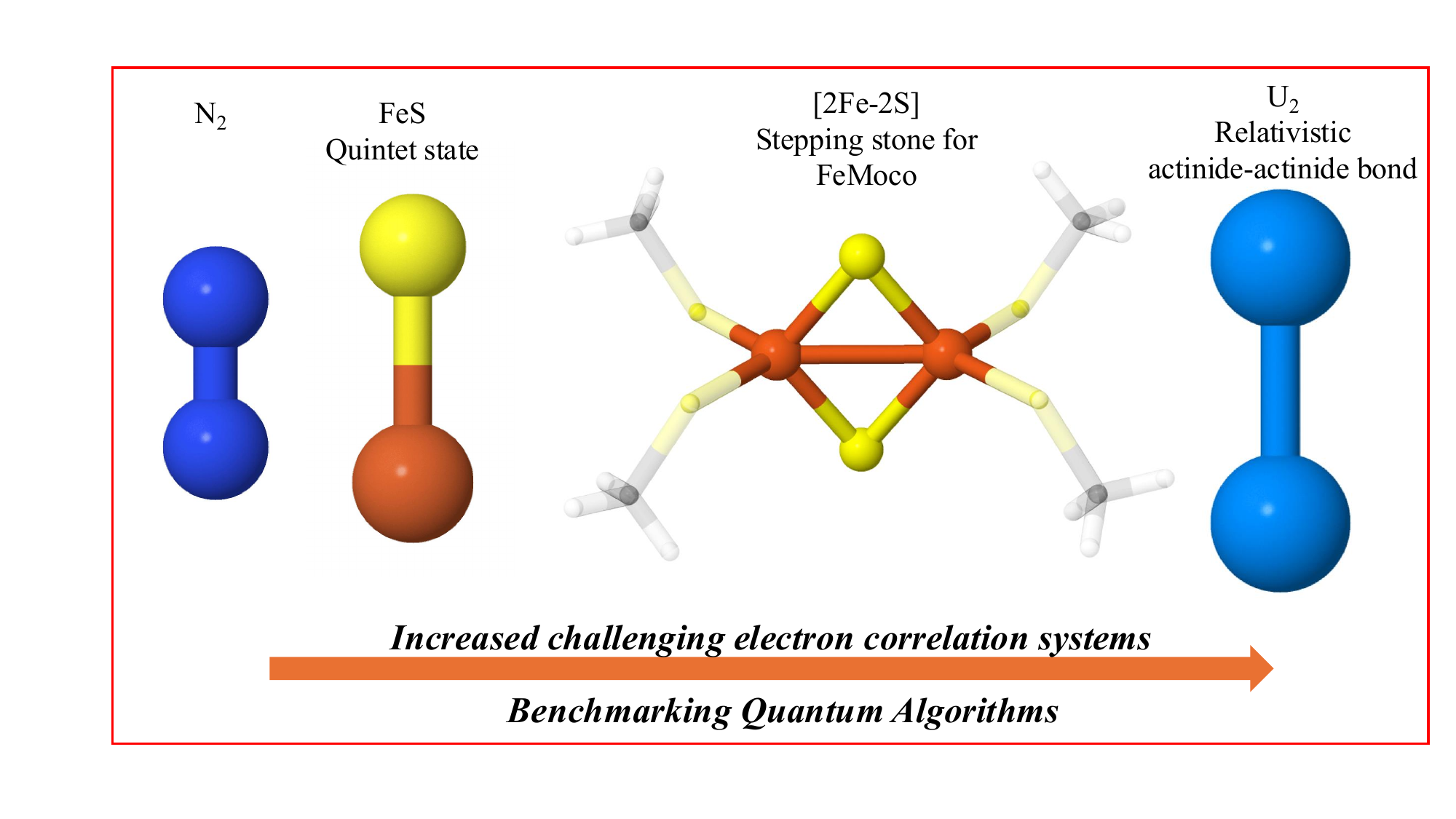}
    \caption{Representative image} %\bp{Bo: this 3D contour seems coming out of nowhere. Should it also be included in the main text for some system? Also, it would be good to remove the cross signs and simple molecules. Instead, focus more on the selected systems, and briefly mentioned the transition rational of the four systems.}}
    \label{TOC}
\end{figure*}

\end{document}